\documentclass[english]{article}
\usepackage[T1]{fontenc}
\usepackage[latin9]{inputenc}
\usepackage{babel}
\usepackage{graphicx,amsmath,amssymb,color}
\usepackage[normalem]{ulem}
\usepackage{amsfonts}
\usepackage[toc,page]{appendix}
\usepackage{hyperref}
\usepackage{latexsym}
\usepackage{amsfonts}
\usepackage{algpseudocode}
\usepackage{amsthm}
\usepackage{mathrsfs}
\usepackage{color,verbatim}
\usepackage{psfrag}
\usepackage{algpseudocode} % for psuedocode
\usepackage{algorithmicx}
\usepackage{algorithm}

\usepackage{rotating} % to get a sideways table
\usepackage{bbold} % to get identity matrix symbol
\usepackage{multirow}

\usepackage{subcaption} % allows for subfigures

\usepackage[svgnames]{xcolor}
\usepackage{tikz}
\usetikzlibrary{positioning}

\usepackage{algorithm}

\newcommand{\bra}[1]{\langle#1 |}
\newcommand{\ket}[1]{|#1 \rangle}

\newcommand{\ketbra}[2]{\vert #1 \rangle \! \langle #2 \vert}

\newcommand{\sandwich}[3]{\left \langle #1 \middle \vert #2 \middle \vert #3 \right\rangle}

\ifx \hideCommentsAndNotes \undefined 

\newcommand{\nick}[1]{[\textcolor{blue}{Nick: #1}]}
\definecolor{myDarkGreen}{rgb}{0, 0.5, 0} 
\newcommand{\ali}[1]{[\textcolor{myDarkGreen}{Ali: #1}]}
\definecolor{orangish}{rgb}{0.75, 0.25, 0} 
\newcommand{\drew}[1]{[\textcolor{orangish}{Drew: #1}]}
\newcommand{\uchenna}[1]{[\textcolor{magenta}{Uchenna: #1}]} 
\newcommand{\removed}[1]{\textcolor{red}{\sout{#1}}}

\else

\newcommand{\nick}[1]{}
\definecolor{myDarkGreen}{rgb}{0, 0.5, 0} 
\newcommand{\ali}[1]{}
\definecolor{orangish}{rgb}{0.75, 0.25, 0} 
\newcommand{\drew}[1]{}
\newcommand{\uchenna}[1]{} 
\newcommand{\removed}[1]{}

\fi

\title{Zeno Blockade Enabling Photonic Quantum Optimization}
\author{Mohammad-Ali Miri, Uchenna Chukwu, and Nicholas Chancellor}
\date{Quantum Computing Inc (QCi), 5 Marine View Plaza; Hoboken, NJ; 07030 USA \today}

\begin{document}

\maketitle

\abstract{
In this work we explore the potential of implementing an optical quantum optimizer using non-linear optics, specifically using sum-frequency generation and/or two photon absorption. This proposal uses Zeno effects to enforce independence constraints and then a linear protocol to find a maximum independent set in a way where the elements of the set can be weighted. Our proposal can either be viewed as an implementation of the entropy computing paradigm presented in [Nguyen et.~al.~Communications Physics 1, 411, 8] which uses real rather than imaginary time evolution, or as quantum annealing within a Zeno constrained subspace. We discuss how such a device could be built, and considerations such as error mitigation, particularly for photon-loss errors. We numerically study aspects of the protocol, including the effect of coherent versus incoherent incarnations of the Zeno effect, finding superior performance from the latter.

\tableofcontents

\section{Introduction}

While universal, fault-tolerant quantum computing provides one vision for the use of quantum devices to compute\cite{Knill2001KLM},  special purpose devices designed to leverage quantum mechanics to solve specific problems also provide a promising route toward useful quantum computing.  One area which is particularly promising for such devices is combinatorial optimization and related sampling problems \cite{Au-Young2023QuantumOpt,Abbas2024QOptReview}. Given the ubiquity of `hard' combinatorial optimization problems in the real world, and their difficulty to solve with classical methods \cite{Rardin1998OpResearch}, these problems are an appealing target.  In this manuscript we present a proposal to build an optical solver for these problems.

By far the most established category of special purpose quantum optimizers are quantum annealers \cite{Finilla1994Qanneal,kadowaki98a,crosson2020prospects,Yarkoni2022AnnealingIndustry,Kendon2026QACMP}.  These devices use quantum evolution, sometimes in conjunction with thermal dissipation effects to solve hard combinatorial optimization problems. Historically,  superconducting flux qubits have been successful as quantum annealers. These devices have historically operated in a regime where thermal equilibration dominates over coherent interference effects.  A regime where the physics is well described by the Kibble-Zurek mechanism in which the system is in thermal equilibrium until the dynamics effectively freeze\cite{Kendon2026QACMP,Kibble76a,Zurek96a,chancellor16b,Bando20a}.  While open system effects in the form of thermal equilibration are highly important for these devices, the equilibrium is fundamentally quantum,  so it is appropriate to treat them as quantum solvers.  In fact recently flux qubit quantum annealers have been demonstrated to have a scaling advantage over state-of-the-art classical methods for approximate optimization. on problems matching the hardware graph.  Engineering improvements in flux qubit devices have allowed such devices to operate in a `diabatic' regime\cite{crosson2020prospects,King2022Coherent,Kendon2026QACMP},  charactorized by coherent interference and no longer well described by the Kibble-Zurek mechanism.  In parallel,  trapped atom technologies \cite{Ebadi2022RydbergAnneal,Goswami2024RydbergQA} have progressed to the point where they can also be used as an annealing platform which also operates in the diabatic regime.

Both flux-qubit and trapped atom technologies however are constrained to limited hardware graphs, in other words they do not natively encode arbitrary interactivity between qubits.  From a purely theoretical perspective, as long as enough interactivity is present, it is always possible to map arbitrary problem connectivity to these devices using techniques such as minor embedding \cite{choi08a,choi10a} or parity based encodings\cite{Lechner15a,Rocchetto16a,Palacios2025Encode}. From a practical perspective however,  such mappings lead to substantial overheads \cite{Kendon2026QACMP}. From general graph-theoretical arguments, the number of qubits in a quasi-2D arrangement to map a highly connected problem must be the square of the number of original variables.  Moreover,  with such mappings, a single qubit flip in the original problem now requires many physical qubits to be flipped, slowing down the effective dynamics. While there has been a recent proposal to implement encoded operations, which would flip all qubits simultaneously \cite{Headley2025Gadgets},  such implementations would come with their own challenges.

One of the key advantages of the optical protocol we propose here is that it naturally allows arbitrary connectivity, and thus avoids overheads from problem mappings. This is a key step toward practical quantum optimizers, recall that while an advantage has been shown in annealing for approximate optimization, this was not for real problems but for artificial ones which natively map to the hardware.  Here we present a route toward a solver capable of approaching arbitrary problems without overhead, by using photons which are naturally mobile and therefore more conducive to high connectivity. The protocol we propose here can be viewed in two ways, it can either be seen as an optical implementation of (Trotterized) quantum annealing within a Zeno-constrained subspace, or as an implementation of the entropy computing \cite{nguyen2025entropycomputing} paradigm based on real-time rather than imaginary-time evolution. 

Aside from annealing, there are a number of other quantum and optical paradigms for special purpose optimizers. For example,  Gaussian Boson sampling can be used in a variational way to solve complex optimization problems\cite{bradler2021ORCA,goldsmith2024ORCA_TSP}. Unlike the proposal here and the annealing paradigm, such techniques do not map the problem directly and therefore rely on machine learning techniques to effectively capture the correlations.  It is worth noting that the protocol we propose here could also have some variational elements, if the control parameters were treated variationally, as is done in the quantum approximate optimization algorithm (QAOA)\cite{Farhi14a,Hadfield2019QAOA}, but this would still be the use of variational techniques on top of a direct problem mapping, so different from variational Boson sampling techniques. The results here are closer to approximate quantum annealing (AQA) as proposed in \cite{Willsch2022AQA}. Coherent Ising machines and early implementations of the entropy computing paradigm do explicitly map problems, but the current large-scale implementations of both involve measurement and feedback in a way which disrupts coherent superpositions \cite{Yamamoto2020CoherentIsing,nguyen2025entropycomputing}. While there are theoretically ways to implement coherent Ising machines in an all-optical way, photons from interference still leave the system carrying information about the logical state in an uncontrolled way which is likely to lead to decoherence\cite{chukwu2025opticalQuantum}.  

The work presented here effectively sets entropy computing apart from coherent Ising machines by presenting a theoretical implementation of entropy computing which is unambiguously fully quantum, an analog of which does not exist for coherent Ising machines. It is worth noting that the proposal here is likely not the only way to implement entropy computing in a fully quantum way.  Moreover, as we discuss later, our protocol admits a natural error mitigation strategy to mitigate against errors caused by photon loss.  We also show how the protocol proposed here can be implemented using a temporal encoding with only two physical nonlinear elements.

The key aspect of the protocol here is the use of quantum Zeno effects to enforce constraints. These constraints appear both in the form of linear constraints necessary to encode an optimization problem (maximum independent set in the discussion here),  and to maintain the qubit subspace. Quantum Zeno effects are the broad class of effects where a system being effectively `watched' can prevent unwanted dynamics\cite{Misra1977Zeno,Itano2009ZenoPerspect}. While originally conceived in terms of quantum measurement, Zeno effects generalize to all  kinds of interactions, including coherent interactions which yield phases.  We examine the effect of both coherent Zeno effects and incoherent Zeno effects in the form of photon loss leading to effective two-photon absorption. These are coupled with linear gain, which has been shown to be desirable in practice\cite{Berwald2025Zeno}. While open system effects such as photon loss can be viewed as unwanted noise, it has been shown that loss base Zeno effects can lead to a theoretical quantum advantage on unstructured search, yielding an analog equivalent of the speedup from Grover's famous algorithm \cite{Grover1997Search,Berwald2024Zeno}. 

The method for maintaining a qubit subspace which we use here can be viewed as an implementation of Zeno blockade to block more than a single photon from entering a mode.  Zeno blockade effects have been discussed theoretically and demonstrated experimentally as a tool for a number of purposes\cite{Huang2010switching,Huang2012entangled,McCusker2013switching}, including nonlinear optics based gates \cite{Sun2013ZenoGates}. They also have been shown to be effective even at the single-photon level \cite{Huang2012Fredkin,Chen2017ZenoChip,Ma2020UltraBrightSource,Ma2023FewPhoton}.

\section{Background}

Before discussing the implementations, it is worth defining some important physical operations which will be used many times, as well as introducing the computational problem which will underpin this work, the maximum independent set problem including two instances used as examples in this study. The main goal of this section is to make the results more accessible to a wider audience and to provide concrete definitions to make the discussion more grounded. This section does not contain any original results.

\subsection{Two-Photon Absorption}

The first of these is the two-photon absorption process (TPA); this process can be defined by applying the following Lindbladian superoperator (acting on mode $j$ We now explicitly talk about how to do this in time bins and give an architecture, it requires some fast switching which would be very difficult in practice I think, but seems like a reasonable starting place.}) to the density matrix
\begin{equation}
    \mathcal{L}^{(j)}_\mathrm{TPA}[\hat{\rho}]=\left[\hat{a}_j\hat{a}_j\hat{\rho} \hat{a}_j^\dagger\hat{a}_j^\dagger-\frac{1}{2}\left(\hat{a}_j^\dagger\hat{a}_j^\dagger\hat{a}_j\hat{a}_j\hat{\rho}+\hat{\rho}\hat{a}_j^\dagger\hat{a}_j^\dagger\hat{a}_j\hat{a}_j \right) \right].
    \label{eq:tpa_lind}
\end{equation}
The action of this term is to incoherently remove two photons from the system. For states with zero or one photons such a process is not possible, so this term does nothing. This can be seen mathematically by the fact that in both cases each term either operates $\hat{a}$ twice from the left or $\hat{a}^\dagger$ twice from the right, or both. For either zero or one photon states this will result in annihilating the vacuum and therefore a zero result. For the two photon state it will remove both photons $\ketbra{2}{2}\rightarrow \ketbra{0}{0}$.

For any time-independent $\mathcal{G}$ which represents a linear operation we can formally solve the differential equation
\begin{equation}
\frac{\partial \hat{\rho}}{\partial t}=\mathcal{G}[\hat{\rho}]
\end{equation}
to obtain
\begin{equation}
\hat{\rho}(t)=\Omega(t)\left[\hat{\rho}\right]=\exp \left[ t\mathcal{G}\right][\hat{\rho}] \label{eq:gen_supOp_sol}
\end{equation}
where exponentiation of the superoperator can be thought of as expanding the Taylor series of the exponential operator in numbers of applications,
\begin{equation}
    \exp \left[ t\mathcal{G}\right][\hat{\rho}]= \sum^\infty_{n=0}\frac{1}{n!}t^n\mathcal{G}[..._n[\hat{\rho}]..._n]
    \label{eq:supOp_exp}
\end{equation}
where for example $\mathcal{G}[..._0\hat{\rho}..._0]=\hat{\rho}$, $\mathcal{G}[..._1\hat{\rho}..._1]=\mathcal{G}[\hat{\rho}]$, $\mathcal{G}[..._2\hat{\rho}..._2]=\mathcal{G}[\mathcal{G}[\hat{\rho}]]$ etc...
In practice, due to the linearity of quantum theory, this formal solution can be readily accessed numerically by representing $t\mathcal{G}$ as a matrix acting on a vectorized version of $\hat{\rho}$ and using standard numerical matrix exponentiation routines. 

In the specific case where $\mathcal{G}=\mathcal{L}^{(j)}_\mathrm{TPA}$ the solution following equation \ref{eq:gen_supOp_sol} corresponds to

\begin{equation}
\hat{\rho}(t)=\Omega^{(j)}_\mathrm{TPA}(t)\left[\hat{\rho}\right]=\exp \left[ t\mathcal{L}^{(j)}_\mathrm{TPA}\right][\hat{\rho}].
\end{equation}

\subsection{Sum Frequency Generation and Coherent Operation \label{sub:sfg_coh}}

We will primarily focus on the difference between loss-based Zeno effects and phase-based Zeno effects, two photon loss can be described by the Lidbladian in eq.~\ref{eq:tpa_lind} which when acting alone, generates the superoperator in eq.~\ref{eq:tpa_drive}. The analogous coherent process is sum frequency generation a process described by the Hamiltonian
\begin{equation}
    \hat{H}_\mathrm{SFG}=\hat{a}_j\hat{a}_j\hat{a}^\dagger_p+\hat{a}^\dagger_j\hat{a}^\dagger_j\hat{a}_p
    \label{eq:H_sfg}
\end{equation}
where mode $p$ is a ``pump'' mode which is generated by the nonlinear process, but unlike in the case of two-photon absorption, will remain available to interact with the original mode. The generator for the coherent process can therefore be written as 
\begin{equation}
    \mathcal{G}_\mathrm{SFG}[\hat{\rho}]=i\hat{\rho}\hat{H}_\mathrm{SFG}-i\hat{H}_\mathrm{SFG}\hat{\rho}.
    \label{eq:G_sfg}
\end{equation}
The process of sum-frequency generation in isolation starting with an empty pump mode and tracing out the pump mode (physically allowing it to leave the system unmeasured) at the end can therefore be represented as applying
\begin{equation}
    \Omega_\mathrm{SFG}(\gamma t)[\hat{\rho}]=\mathrm{Tr}_p\left[\exp\left[\gamma t \mathcal{G}_\mathrm{SFG}\right]\left[\hat{\rho}\otimes\ketbra{0_p}{0_p} \right]\right].
    \label{eq:Omega_sfg}
\end{equation}
where the exponentiation is defined the same as in equation \ref{eq:supOp_exp} and the partial trace over the pump mode is defined as
\begin{equation}
    \mathrm{Tr}_p\left[\hat{\rho}\right]=\sum^{\infty}_{j=0}\sandwich{j_p}{\hat{\rho}}{j_p}
\end{equation}
where the tensor product with other modes is implied so the result is a matrix rather than a single number. Because the pump mode may not be empty when traced over, the total action of of the sum-frequency-generation superoperator defined in equation \ref{eq:Omega_sfg} does not necessarily act coherently, if we assume that the system is in the $\{\ket{0},\ket{1},\ket{2}\}$ subspace, than performing sum frequency generation for a time such that $\gamma t=\frac{\pi}{4\sqrt{2}}$ would result in complete loss of the two photon state $\ket{2}$, and have no effect on the other two states. This is identical to the effect of strong two-photon absorption. On the other hand if $\gamma t=\frac{\pi}{2\sqrt{2}}$, than the pump mode photon will be able to return, effectively leading to a coherent operation which sends $\ket{2}\rightarrow-\ket{2}$. This is effectively the ``phase flip'' operation discussed in \cite{Berwald2024Zeno}, which was shown in that work to be able to support a Zeno effect capable of achieving an optimal quantum speedup. Additional phases can also be applied to the pump mode, a fact which will prove to be important in the protocols we develop later. Intermediate values will lead to intermediate behavior. As we explore later in this manuscript, this provides a convenient method to explore the interpolation between fully coherent operation and fully incoherent operation when enforcing the constraints.

\subsection{Maximum Independent Set}

The computational problem we will primarily focus on solving in this manuscript is the maximum independent set problem and its weighted variant. The problem is, given an undirected graph $G$, with vertices $V$ and edges $E$, what is the maximum number of vertices which can be selected such that none share an edge. This problem can be expressed as a linear programming problem
\begin{align}
    \mathrm{ maximise} \sum_i s_i \nonumber \\
    s_i+s_j\le 1 \forall \{i,j\} \in E \nonumber \\
    s_i\in\{0,1\} \forall i,
\end{align}
and can be generalized to a weighted variant, where each node is assigned a weight $w_i>0$
\begin{align}
    \mathrm{ maximise} \sum_i w_is_i \nonumber \\
    s_i+s_j\le 1 \forall \{i,j\} \in E \nonumber \\
    s_i\in\{0,1\} \forall i. \label{eq:wmis_def}
\end{align}
The maximum independent set problem and by extension the weighted variant is known to be NP-hard, in fact it can be mapped in a simple one-to-one reduction to set packing\cite{Neuwohner2021MISsetPack}, one of Karp's \cite{Karp1972Reduce} original $21$ NP-complete problems. Since the NP-hardness of (weighted) maximum independent set implies that all other hard optimization problems can be reduced to it, and it is even hard to find approximate solutions \cite{Berman1995MIShardApprox} it is a natural target for quantum and unconventional optimizers. Rydberg atom based quantum annealers for example have a natural mapping to these problems, and it has been explored heavily in that context \cite{Ebadi2022RydbergAnneal,Hyeonjun2024RydbergAnneal,deOliveira2025RydbergAnneal}. It is worth noting that Rydberg systems, similar to superconducting annealers, require gadgets to achieve effective full connectivity, leading to experimental overheads \cite{Xingze2020RydbergEmbed,Nguyen2023RydbergEmbed,Cazals2025RydbergGadgets}.

\begin{figure}
    \centering
    \includegraphics[width=0.35\linewidth]{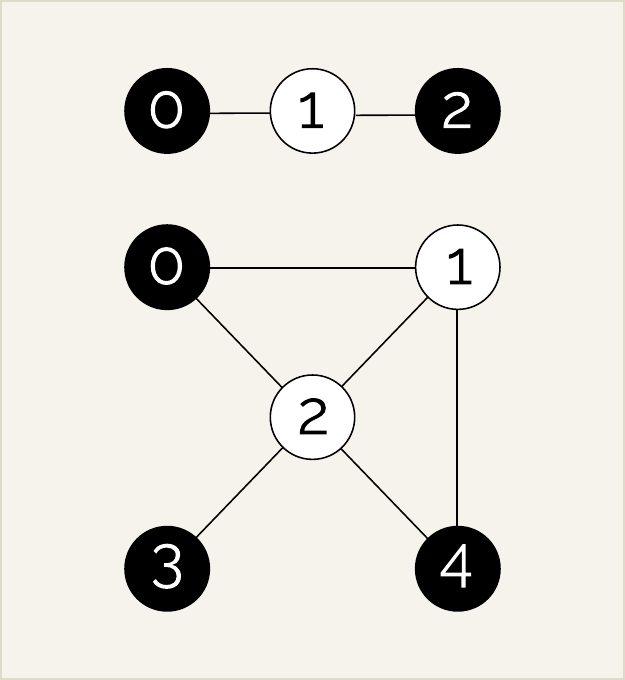}
    \caption{Graphs used for maximum independent set examples in this problem with the MIS highlighted. The top graph is the one which appears in equation \ref{eq:three_node_line}, and the bottom in equation \ref{eq:five_node_graph}.}
    \label{fig:MIS_graphs}
\end{figure}

For the purpose of this paper we will study the dynamics of solving small maximum independent set problems. A simple problem we will examine is the finding the maximum independent set on a three-node line, in other words on the graph where 
\begin{align}
   V=\{0,1,2\} \nonumber \\
   E=\{\{0,1\},\{1,2\}\}. \label{eq:three_node_line}
\end{align}
By inspection (or exhaustive search), it can easily be shown that the set $s=\{0,2\}$ is the unique maximum independent set on this graph. We also use a graph with five nodes in some examples 
\begin{align}
   V=\{0,1,2,3,4\} \nonumber \\
   E=\{\{0,1\},\{0,2\},\{1,2\},\{1,4\},\{2,3\},\{2,4\}\}. \label{eq:five_node_graph}
\end{align}
The maximum independent set for this graph can again be found by inspection and verified exhaustively, it is $s=\{0,3,4\}$. Both graphs are visualized in figure \ref{fig:MIS_graphs}.

\section{Subspace Confinement \label{sec:subspace_confine}}

We describe a number of ways to use Zeno effects to implement quantum annealing over maximum independent set problems (and the weighted variant in appendix \ref{sub:wmis}). Since such problems are NP-hard, this implies that all NP-hard optimization problems (which will necessarily have NP-complete decision variants) can be mapped to these devices \cite{Cook1971NPHard,Karp1972Reduce}. 

We consider a conceptually simple encoding of a qubit based on the presence or absence of a single photon in a single mode. In this, so-called single-rail, encoding the presence of a single photon corresponds to a logical $\ket{1}$ state and the absence of any photons corresponds to the $\ket{0}$ state. All other states do not have a logical meaning and correspond to a failure of the algorithm.

\subsection{Zeno Blockade and Driving\label{sub:ZB_drive}}

\begin{figure}[h]
\centering
\begin{tikzpicture}[>=stealth,thick,node distance=1.6cm]

    %Full Fock Space box
    \node[draw, rounded corners, inner sep=6pt, align=center] (full) {Higher states\\[3pt] $\ket{2}\;\;\ket{3}\;\;\ket{4}\;\dots$};

    %Computational Subspace box
    \node[draw, rounded corners, inner sep=6pt, align=center, below=2.0cm of full] (comp){Computational Subspace\\[3pt] $\{\ket{0},\,\ket{1}\}$};

    % Right: attempted excitation (dashed up)
    \draw[->, dashed] ([xshift=0.7cm]comp.north) -- 
    node[left,align=left] {\small Blockaded Linear \\[-2pt]\small displacement to $\ket{2}$\\[-2pt]\small during driving}
    ([xshift=0.7cm]full.south);

    % Right: Sum frequency Generation (solid)
    \draw[->] ([ yshift=0 cm]full.east) .. controls +(0.8,-0.7) and +(0.8,0.7) .. node[right,align=left] {\small Phase acquired \\[-2pt]\small during sum-frequency \\[-2pt]\small generation}([yshift=0.1 cm]full.east);

    % Right: rapid decay back down (solid)
    \draw[->] ([xshift=1.15cm]full.south) .. controls +(0.8,-0.7) and +(0.8,0.7) .. node[right,align=left] {\small Rapid decay via\\[-2pt]\small two-photon absorption }([xshift=1.15cm]comp.north);

    % Right: Desired driving
    \draw[->] ([ yshift=0cm]comp.east) .. controls +(0.8,-0.7) and +(0.8,0.7) .. node[right,align=left] {\small Desired driving}([yshift=0.1 cm]comp.east);

    \end{tikzpicture}
    \caption{ Conceptual diagram of subspace confinement and driving implemented using a combination of coherent  (sum frequency generation) and incoherent (two-photon absorption or equivalent.) processes.}
    \label{fig:qa_two_photon_loss}
\end{figure}
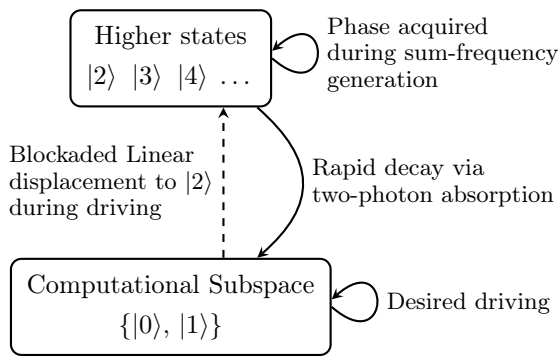

Given our encoding, the first consideration is how to prevent more than one photon from being present in a mode. To do this, we can implement displacement on a mode (for example by weakly beamsplitting with a strong coherent state), in the presence of strong two-photon absorption. The density operator would therefore obey the differential equation:
\begin{equation}
    \frac{\partial \hat{\rho}}{\partial t}=c\mathcal{G}_\mathrm{disp}\left[\hat{\rho}\right]+\gamma \mathcal{L}_\mathrm{TPA}[\hat{\rho}]
    \label{eq:tpa_drive}
\end{equation}
where we have arbitrarily fixed the phase of the displacement which is performed without loss of generality. The Lindbladian $\mathcal{L}_\mathrm{TPA}$ is defined in equation \ref{eq:tpa_lind}.  We further define the generator for displacement as
\begin{equation}
    \mathcal{G}_\mathrm{disp}\left[\hat{\rho}\right]=\left(i\hat{\rho}\left(\hat{a}+\hat{a}^\dagger \right)-i\left(\hat{a}+\hat{a}^\dagger \right)\hat{\rho}\right). \label{eq:G_disp}
\end{equation}
If we now consider the case where $\frac{\gamma}{c}\gg 1$ and consider starting in the vacuum, than a single photon can be added since  $\mathcal{L}_\mathrm{TPA}[\hat{\rho}]$ does nothing in the $\{\ket{0},\ket{1}\}$ subspace.  However adding a photon when one is already present would result in the $\ket{2}$ state, which would decay very quickly to $\ket{0}$ therefore, a Zeno effect will be present and will restrict to the subspace making the evolution approximately
\begin{align}
    \frac{\partial \hat{\rho}}{\partial t}\approx \nonumber \\ ic\left(\hat{\rho}\left(\ketbra{0}{1}+\ketbra{1}{0} \right)-\left(\ketbra{0}{1}+\ketbra{1}{0} \right)\hat{\rho}\right)=ic\left(\hat{\rho}\hat{X}-\hat{X}\hat{\rho}\right)=c\mathcal{G}_X\left[\hat{\rho}\right]
    \label{eq:tpa_drive_strong_zeno}
\end{align}
Where $\hat{X}$ is the Pauli-X operation. 
This equation is effectively coherent flipping between the zero and one photon state, and in the limit of a strong Zeno effect operates entirely coherently, as seen by the fact that the evolution is effectively described by a von Neumann equation with no additional open-system terms. Formally solving equation \ref{eq:tpa_drive_strong_zeno} yields a unitary driving operation of the form
\begin{equation}
    \Omega_\mathrm{drive}(c)\left[\hat{\rho}\right]\approx \exp\left(c\mathcal{G}_{X}\right)\left[\hat{\rho}\right].
\end{equation}
A diagram showing how this works conceptually and how it can be extended to coherent operation by sum-frequency generation appears in figure \ref{fig:qa_two_photon_loss}.

\begin{figure}
    \centering
    \includegraphics[width=0.75\linewidth]{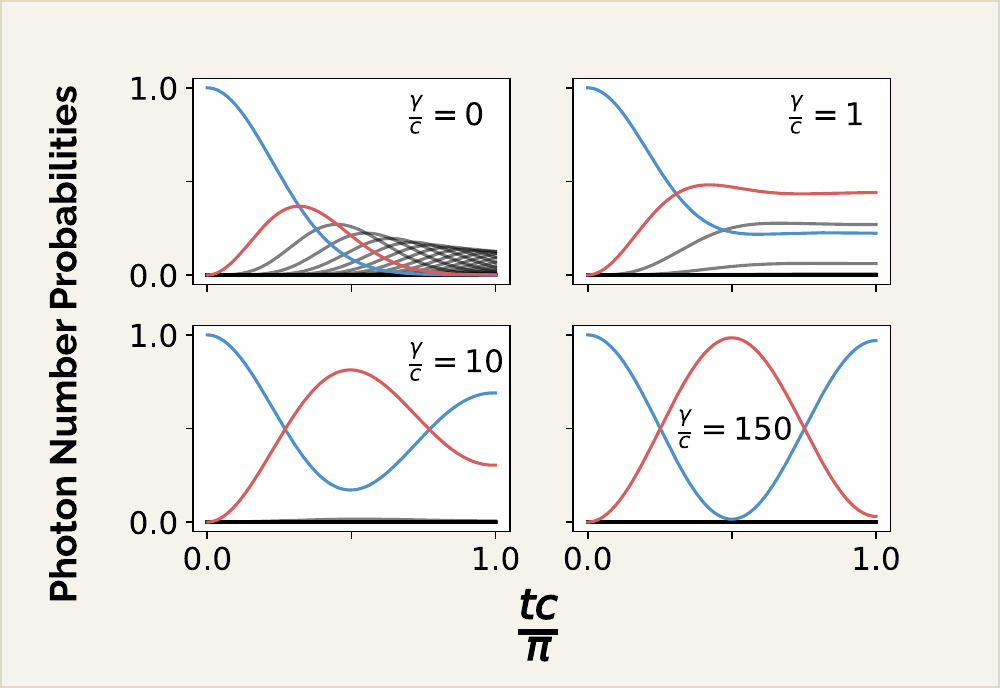}
    \caption{Onset of Zeno blockade from two-photon absorption which restricts the system to the $\{\ket{0},\ket{1}\}$ subspace. The dynamics will obey equation \ref{eq:tpa_drive}, with the ratio $\frac{\gamma}{c}$ given on each plot. The blue line is probability to be in $\ket{0}$, the red is probability to be in $\ket{1}$, and the probability for all higher photon number states are represented as semi-transparent black lines. The Hilbert space is limited to a maximum of $30$ photons.}
    \label{fig:Zeno_onset_TPA}
\end{figure}

The approach to this behavior is visualized in figure \ref{fig:Zeno_onset_TPA}. With no two-photon absorption the average photon number grows quickly leaving both the $\ket{0}$ and $\ket{1}$ state almost empty. For a moderate value the displacement and absorption quickly reach an equilibrium balance. For absorption which is ten times as strong as displacement, the system is well confined in $\{\ket{0},\ket{1}\}$ but there is still substantial decoherence. Finally, when the absorption is $150$ times as strong, we see a nearly perfectly coherent bit flipping operation within this subspace.

If we consider the driving as being performed by weakly coupling to a strong coherent state, than our driving is an example of Zeno blockade, a single photon is allowed to enter, but a second one is not \cite{Sun2013ZenoGates}. 

\begin{figure}
    \centering
    \includegraphics[width=0.75\linewidth]{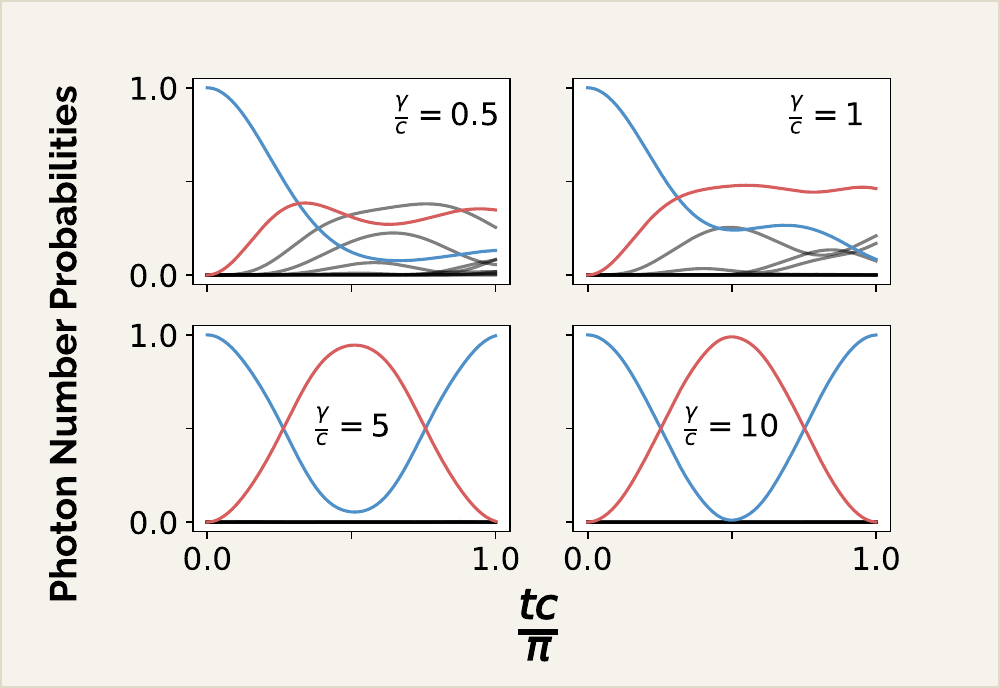}
    \caption{Onset of Zeno blockade from coherent sum-frequency generation which restricts the system to the $\{\ket{0},\ket{1}\}$ subspace. The dynamics are those which result from applying equation \ref{eq:sfg_drive_supOp} to the vacuum, with the ratio $\frac{\gamma}{c}$ given on each plot. The blue line is probability to be in $\ket{0}$, the red is probability to be in $\ket{1}$, and the probability for all higher photon number states are represented as semi-transparent black lines. The Hilbert space is limited to a maximum of $10$ photons. Note that the $\gamma=0$ case is identical to that shown in figure \ref{fig:Zeno_onset_TPA}}
    \label{fig:Zeno_onset_SFG}
\end{figure}

It is possible to also achieve the same effect based on coherent sum frequency generation, in this case instead of the time evolution obeying equation \ref{eq:tpa_drive}, it would obey
\begin{equation}
    \frac{\partial \hat{\rho}}{\partial t}=c\mathcal{G}_\mathrm{disp}\left[\hat{\rho}\right]+\gamma \mathcal{G}_\mathrm{SFG}\left[\hat{\rho}\right] \label{eq:sfc_drive}
\end{equation}
where $\mathcal{G}_\mathrm{disp}\left[\hat{\rho}\right]$ is defined in equation \ref{eq:G_disp} and $\mathcal{G}_\mathrm{SFG}\left[\hat{\rho}\right]$ is defined in equation \ref{eq:G_sfg}. While the time evolution defined in equation \ref{eq:sfc_drive} is completely coherent, there is still a route to induce decoherence since the pump mode must be traced over at the end of the application. The overall superoperator for the driving is therefore
\begin{equation}
    \Omega_\mathrm{d,SFC}(c,\gamma,t)\left[\hat\rho\right]=\mathrm{Tr}_p\left[\exp\left( tc\mathcal{G}_\mathrm{disp}+t\gamma \mathcal{G}_\mathrm{SFG}
    \right)\left[\hat{\rho}\otimes\ketbra{0_p}{0_p}\right] \right]. \label{eq:sfg_drive_supOp}
\end{equation}
In the limit where $\frac{\gamma}{c}\gg 1$, there will effectively be a coherent Zeno effect from phase reversal and the action of this superoperator will effectively be to flip between $\ket{0}$ and $\ket{1}$. As with two-photon absorption, we can observe the approach to the Zeno blockade behaviour, this is depicted in figure \ref{fig:Zeno_onset_SFG}. We note that this happens at a much lower value of $\frac{\gamma}{c}$ than for two-photon absorption, with the bit flipping being visually indistinguishable from perfect operation for $\frac{\gamma}{c}=10$. This behavior is more fully explored in section \ref{sub:single_mode_drive}.

\begin{figure}
    \centering
    \includegraphics[width=0.5\linewidth]{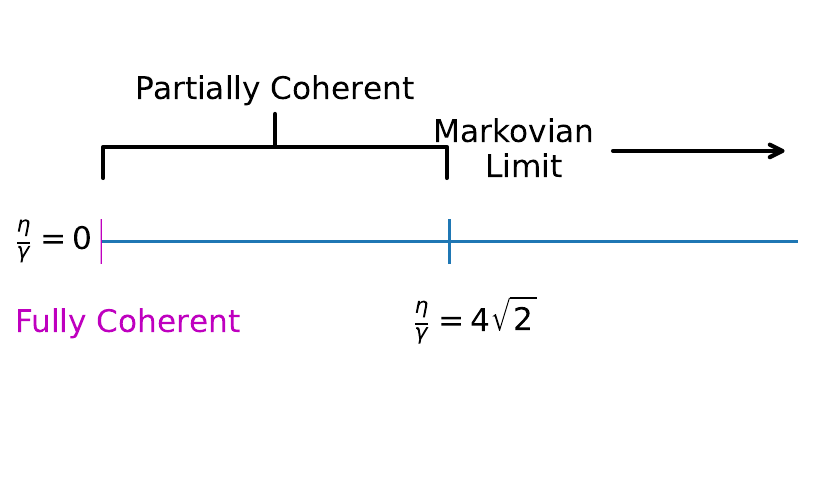}
    \caption{Illustrations of the regimes of dynamics for lossy sum-frequency generation described by  $\Omega_\mathrm{dl,SFG}$ from equation \ref{eq:Omega_dl_SFG}. The operation is fully coherent and effectively unitary only at $\frac{\eta}{\gamma}=0$, but allows for partially coherent return of amplitude until critical damping is achieved at$\frac{\eta}{\gamma}=4\sqrt{2}$. This approach is illustrated in figure \ref{fig:rho_12_lossy_sfg}.  Even at this point the dependence of the loss rate on the pump mode occupation makes the superoperator act in a non-Markovian way, as loss dominates, the dynamics approach the limit of Markovian two-photon absorption, as visualized in figure \ref{fig:rho_12_markov_approach} (appendix \ref{app:lossy_SFG_coherence}).}
    \label{fig:regime_cartoon}
\end{figure}

As with the coherent constraints, it is possible to interpolate between fully incoherent and a fully coherent Zeno effects (as the IQZ and CQZ regimes discussed in \cite{Sun2013ZenoGates}) by adding single photon loss to the pump mode. The Lindbladian for single-photon loss is 
\begin{equation}
\mathcal{L}^{(j)}_\mathrm{loss}[\hat{\rho}]=\hat{a}_j\hat{\rho} \hat{a}_j^\dagger-\frac{1}{2}\left(\hat{a}_j^\dagger\hat{a}_j\hat{\rho}+\hat{\rho}\hat{a}_j^\dagger\hat{a}_j \right) .
    \label{eq:loss_lind}
\end{equation}
making use of this Lindbladian, we define 
\begin{align}
    \Omega_\mathrm{dl,SFG}(c,\gamma,\eta,t)\left[\hat\rho\right]=\nonumber \\
    \mathrm{Tr}_p\left[\exp\left( tc\mathcal{G}_\mathrm{disp}+t\gamma \mathcal{G}_\mathrm{SFG}+t\eta\mathcal{L}^{(p)}_\mathrm{loss}
    \right)\left[\hat{\rho}\otimes\ketbra{0_p}{0_p}\right] \right]. \label{eq:Omega_dl_SFG}
\end{align}
as we show in appendix \ref{app:lossy_SFG_coherence}, the coherence between the two-photon and loss mode obeys the differential equation for a damped harmonic oscillator, at or beyond the critical damping point $\eta\ge 4 \sqrt{2}\gamma$. An interpolation between coherent and incoherent Zeno blockade can be performed between $\eta=0$ (fully coherent), and $\eta=4\sqrt{2}$ (fully incoherent). While conceptually similar to two photon absorption, this fully incoherent limit isn't fully equivalent to the two-photon absorption defined in equation \ref{eq:tpa_lind} as the occupation of the pump mode will make it act in a non-Markovian way. The Markovian limit is approached as $\eta$ becomes very large, and this approach is illustrated in appendix \ref{app:lossy_SFG_coherence}. The regimes of operation are depicted visually in figure \ref{fig:regime_cartoon}. We show later that while there is a large difference between the incoherent and coherent manifestation of the Zeno effect, there is only a small difference between the Markovian and non-Markovian loss in the incoherent regime. 
\begin{figure}
    \centering
    \includegraphics[width=0.5\linewidth]{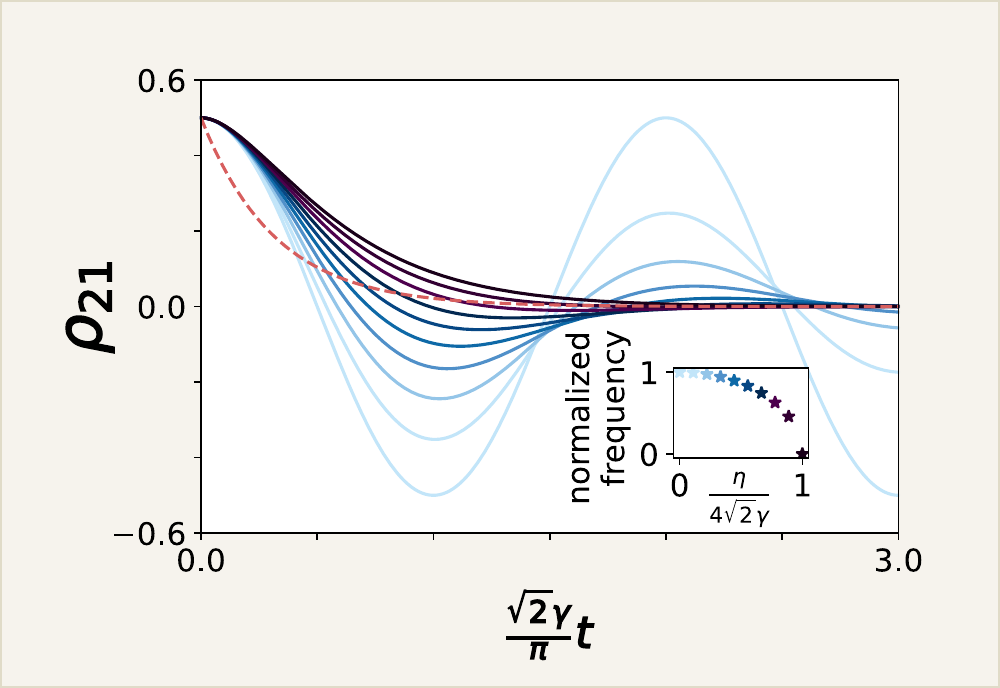}
    \caption{Plot of $\rho_{12}=\sandwich{1,0_p}{\hat{\rho}}{2,0_p}$ versus time for time evolution with the superoperator $\Omega_\mathrm{dl,SFG}(c=0,\gamma,\eta,t)\left[\hat{\rho}_0\right]$ from equation \ref{eq:Omega_dl_SFG}. The system is initialized in $\hat{\rho}_0=\frac{1}{2}(\ket{1}+\ket{2})(\bra{1}+\bra{2})$ with the pump mode empty for $c=0$ and varying values of $\frac{\eta}{\gamma}$ between undamped ($\eta=0$) and critically damped ($\eta=4\sqrt{2}\gamma$). The inset shows the frequency for each value of $\eta$ (using the same color scheme) and normalized to the frequency with no photon loss. The red dashed line shows two-photon absorption with a loss rate of $\gamma$ for comparison.}
    \label{fig:rho_12_lossy_sfg}
\end{figure}
Figure  \ref{fig:rho_12_lossy_sfg} illustrates the transition from coherent return of an off diagonal coherence element of the density matrix to loss in a simple case where the system is initialized in $\hat{\rho}=\frac{1}{2}(\ket{1}+\ket{2})(\bra{1}+\bra{2})$ with the pump mode empty for $c=0$ and varying values of $\frac{\eta}{\gamma}$ between undamped and critically damped. 

\subsection{Incoherent Constraints\label{sub:HOM_constr}}

To implement the independence constraints necessary for maximum independent set, we exploit the Hong-Ou-Mandel effect, a well known optical effect whereby if two photons enter a $50:50$ beamsplitter each in one of the modes, then they will both exit in the same mode. Mathematically
\begin{equation}
\ket{1}_j\ket{1}_k\rightarrow\frac{1}{\sqrt{2}}\left(\ket{2}_j\ket{0}_k+\ket{0}_j\ket{2}_k\right)
\end{equation}
From this fact we note something interesting, if we apply\footnote{Since operations commute on different modes applying these in either order is the same as applying them simultaneously.} \begin{equation}
    \Omega^{(k)}_\mathrm{TPA}(\gamma t)[\Omega^{(j)}_\mathrm{TPA}(\gamma t)[\hat{\rho}]] \label{eq:two_mode_TPA}
\end{equation}
nothing will happen to any configuration which began with a total number of photons less than two, but if both modes have a photon, that photon will leave the system. If we apply this operation for a time such that $\gamma t \gg 1$, than the state with two photons will be returned to vacuum. 

Since beamsplitting is a unitary operation, it can be inverted, we can therefore construct an operation which sends $\ket{1}_j\ket{1}_k$ to vacuum while leaving all other zero-and-one-photon states alone by applying
\begin{equation}
    \Omega_\mathrm{TPA\, bs}(\gamma)\left[\hat{\rho}\right]= \hat{U}^\dagger_{bs}\Omega^{(k)}_\mathrm{TPA}(\gamma t)\left[\Omega^{(j)}_\mathrm{TPA}(\gamma t)\left[\hat{U}_{bs}\hat{\rho}\hat{U}^\dagger_{bs}\right]\right]\hat{U}_{bs}. \label{eq:bs_TPA_bs}
\end{equation}
To demonstrate this mathematically, we consider what this rotation will do to the generating Lindbladian given in equation \ref{eq:tpa_lind}, in other words to calculate
\begin{equation}
\mathcal{L}^{(j,k)}_{rot}=\hat{U}^\dagger_{bs(j,k)}\mathcal{L}^{(k)}_\mathrm{TPA}(\gamma t)\left[\mathcal{L}^{(j)}_\mathrm{TPA}(\gamma t)\left[\hat{U}_{bs(j,k)}\hat{\rho}\hat{U}^\dagger_{bs(j,k)}\right]\right]\hat{U}_{bs(j,k)}.
\end{equation}
We observe that
\begin{equation}
\hat{U}^\dagger_{bs(j,k)}\hat{a}_j\hat{a}_j\hat{U}_{bs(j,k)}=\frac{1}{2}(\hat{a}_j+i\hat{a}_k)(\hat{a}_j+i\hat{a}_k)=i\hat{a}_j\hat{a}_k+\frac{1}{2}\hat{a}_j\hat{a}_j-\frac{1}{2}\hat{a}_k\hat{a}_k
\end{equation}
we further notice that when restricting to zero or one photon in each mode the second and third terms will have no effect. After applying similar expansions for $\hat{a}_k\hat{a}_k$, $\hat{a}^\dagger_j\hat{a}^\dagger_j$, and $\hat{a}^\dagger_k\hat{a}^\dagger_k$, and grouping together only terms which involve at most single operators in each mode, we have:
\begin{equation}
\mathcal{L}^{(j,k)}_{rot}=\hat{a}_j\hat{a}_k\hat{\rho}\hat{a}^\dagger_j\hat{a}^\dagger_k-\frac{1}{2}\left[\hat{a}^\dagger_j\hat{a}^\dagger_k\hat{a}_j\hat{a}_k\hat{\rho}+\hat{\rho}\hat{a}^\dagger_j\hat{a}^\dagger_k\hat{a}_j\hat{a}_k \right]+\mathcal{G}^{(j,k)}_{\mathrm{high}}\left[\hat{\rho} \right],
\end{equation}
where $\mathcal{G}^{(j,k)}_{\mathrm{high}}\left[\hat{\rho} \right]$ is the parts of the generator made of terms containing higher than single powers of $\hat{a}$ or $\hat{a}^\dagger$ in at least one of the modes, which can be ignored due to the restriction on photon number in each mode. Restricting explicitly to the zero an one photon sector of each mode, we have
\begin{equation}
\mathcal{L}^{(j,k)}_{rot}[\hat{\rho}]\rightarrow \ketbra{0_j0_k}{1_j1_k}\hat{\rho}\ketbra{1_j1_k}{0_j0_k}-\frac{1}{2}\left[\ketbra{1_j1_k}{1_j1_k}\hat{\rho}+\hat{\rho}\ketbra{1_j1_k}{1_j1_k} \right]
\end{equation}
which is exactly the generator for incoherently sending $\ket{1_j1_k}$ to $\ket{0_j0_k}$. 

\subsection{(Partially) Coherent Constraints}

Analogously to the Zeno blockade for driving, it is natural to ask whether phases can be applied to implement constraints via a coherent Zeno effect. As we have discussed in section \ref{sub:sfg_coh}, applying sum-frequency generation can lead to a phase flip in $\ket{2}$ relative to $\{\ket{0},\ket{1} \}$. As we discuss later in this section, this phase shift on its own is not sufficient to implement the constraint. However, a sum -frequency-generation-based system which performs a lesser phase shift can be effective. 

When a full $-1$ phase is applied, it only reverses the direction of the next driving operation, and the following constraint will restore the original direction. If the return is incomplete, this can lead to ``leakage'' out of the constrained subspace. To mitigate this issue, additional phase shifting can be applied to the pump mode before it is returned.  Since sum-frequency generation described by equation \ref{eq:Omega_sfg} run for $\gamma t=\frac{\pi}{4\sqrt{2}}$ converts $\ket{2,0_p}\rightarrow i\ket{0,1_p}$, while doing nothing to $\ket{0,0_p}$ and $\ket{1,0_p}$, it provides us a vehicle to provide our phase, specifically, instead of just applying sum-frequency generation we consider applying
\begin{align}
    \Omega_\mathrm{nl,phase}(\phi_Q,\gamma t=\frac{\pi}{4\sqrt{2}})\left[\hat{\rho}\right] \nonumber \\ =\Omega_{SFG}\left(\frac{\pi}{4\sqrt{2}}\right)\left[\Omega^\mathrm{(p)}_\mathrm{phase}\left(-\phi_Q\right)\left[\Omega_{SFG}\left(\frac{\pi}{4\sqrt{2}}\right)\left[\hat{\rho}\right]\right]\right] \label{eq:Omega_nl_phase}
\end{align}
where $\Omega_{SFG}$ is defined in equation \ref{eq:Omega_sfg}, and $\Omega^\mathrm{(p)}_\mathrm{phase}$ applies a phase to the pump mode:
\begin{equation}
 \Omega^\mathrm{(p)}_\mathrm{phase}\left(\phi\right)\left[\hat{\rho}\right]=\exp\left[-i\phi\hat{a}_p^\dagger\hat{a}_p\right]\hat{\rho}\exp\left[i\phi\hat{a}_p^\dagger\hat{a}_p\right].
\end{equation}
this allows an arbitrary phase to be applied. Various arguments can be made about the maximum phase which can be applied to avoid violating the independence condition for a given graph. One could use a heuristic approach, tuning this value down if the independence condition is frequently violated. It is worth noting that this phase does not have to take it's maximum value for the blockade to be effective. For this reason a conservative bound could also be used. For example we could take
\begin{equation}
    \phi_Q=\pi+\frac{\pi}{d}.
\end{equation}

In general, we will be interested in the case where both loss and phase effects are present, in this case we can combine equation \ref{eq:Omega_nl_phase} with equation \ref{eq:two_mode_TPA} to obtain
\begin{align}
    \Omega_\mathrm{nl}(\phi_Q,\gamma t_\gamma,\eta t_\eta)\left[\hat{\rho}\right]=
    \nonumber \\
    \Omega_{SFG}\left(\gamma t_\gamma\right)\left[\Omega^\mathrm{(p)}_\mathrm{phase,loss}\left(-\phi_Q,\eta t_\eta\right)\left[\Omega_{SFG}\left(\gamma t_\gamma\right)\left[\hat{\rho}\right]\right]\right] \label{eq:Omega_nl_phase_loss}
\end{align}
where
\begin{equation}
    \Omega^\mathrm{(p)}_\mathrm{phase,loss}(\phi_Q,\eta t)=\exp\left(\eta t\mathcal{L}^\mathrm{(p)}_\mathrm{loss}+\phi_Q\mathcal{G}^\mathrm{(p)}_{\mathrm{phase}}\right)\label{eq:Omega_phase_loss}
\end{equation}
 where $\mathcal{G}^\mathrm{(p)}_{\mathrm{phase}}[\hat{\rho}]=-i\hat{a}^\dagger_p\hat{a}_p\hat{\rho}+i\hat{\rho}\hat{a}^\dagger_p\hat{a}_p$.

 Combining with the construction in section \ref{sub:HOM_constr}, a general constraint which can range from fully coherent to fully incoherent with an arbitrary phase can be implemented using

\begin{align}
    \Omega_\mathrm{constr}^{(j,k)}(\phi_Q,\gamma t,\eta t)\left[\hat{\rho}\right]=\nonumber \\
    \hat{U}^{(j,k)\dagger}_{bs}\Omega^{(j)}_\mathrm{nl}(\phi_Q,\gamma t,\eta t)\left[\Omega^{(k)}_\mathrm{nl}(\phi_Q,\gamma t,\eta t)\left[\hat{U}^{(j,k)}_{bs}\hat{\rho}\hat{U}^{(j,k)\dagger}_{bs}\right]\right]\hat{U}^{(j,k)}_{bs}. \label{eq:master_constraint}
\end{align}

Note that partial loss can be achieved by adjusting $\eta t$ or $\gamma t$, so in  mathematical sense these two degrees of freedom are redundant, in this work we elect to interpolate through the whole coherence range by setting $\eta t_\eta=0$ and interpolating $\frac{\pi}{4\sqrt{2}}\le \gamma t_\gamma \le \frac{\pi}{2\sqrt{2}}$.

\section{Optimization Protocol \label{sec:opt_protocol}}

\subsection{Conceptual Schematic}

To make use of an effective Zeno effect from the constraints devised in section \ref{sub:HOM_constr}, we have to slowly increase the photon probability, such that the amplitude to be in $\ket{1_j1_k}$ is never large. To accomplish this, we need to make use of the driving in section \ref{sub:ZB_drive}, but need to use an additional strategy. To do this we first consider a single mode which does not interact with any others, and also consider the possibility of applying a phase shift to the mode. By acting on it with
\begin{equation}
    \hat{U}_\mathrm{phase}(\phi)=\exp\left[-i\phi\hat{a}^\dagger\hat{a}\right].
\end{equation}
Phase shifting does not affect photon number, but does apply a relative phase between photon number states, if we define the phase shifting operation within the zero and one photon subspace, we have
\begin{equation}
\hat{U}_\mathrm{phase}(\phi)=\exp\left[-i\phi\ketbra{1}{1}\right]=\exp\left[i\phi\frac{1}{2}\left(\hat{Z}-\hat{\mathbb{1}}\right)\right],
\end{equation}
where the identity term only contributes a physically meaningless global phase and therefore can be ignored. Note that in practice instead of shifting the phase of the modes, the phase of the coherent states which are used for driving could be shifted each cycle, which is effectively the same operation just performed in a rotating basis. Since we now have Pauli-X driving and Pauli-Z phase terms we can apply a digitized transverse Ising annealing. Effectively, if both $\phi$ and $c$ are small then
\begin{equation}
    \hat{U}_\mathrm{drive}(c)\hat{U}_\mathrm{phase}(\phi)\approx \exp\left(-ic\hat{X}-i\phi \hat{Z}\right).
\end{equation}
if we initially start with $\frac{\phi}{c}\ll -1$, then the ground state of the effective Hamiltonian $c\hat{X}+\phi \hat{Z}$ will to a very good approximation be $\ket{0}$, if we than iteratively apply this operation while slowly reducing the magnitude of $\frac{\phi}{c}$, the system will adiabatically follow its ground state, which in turn will start to contain more and more of the $\ket{1}$ state. When $\phi=0$, the ground state will be an equal superposition. Finally if we ramp to $\frac{\phi}{c}\gg 1$, we will adiabatically follow the ground state to $\ket{1}$.

\begin{figure}
    \centering
    \includegraphics[width=0.5\linewidth]{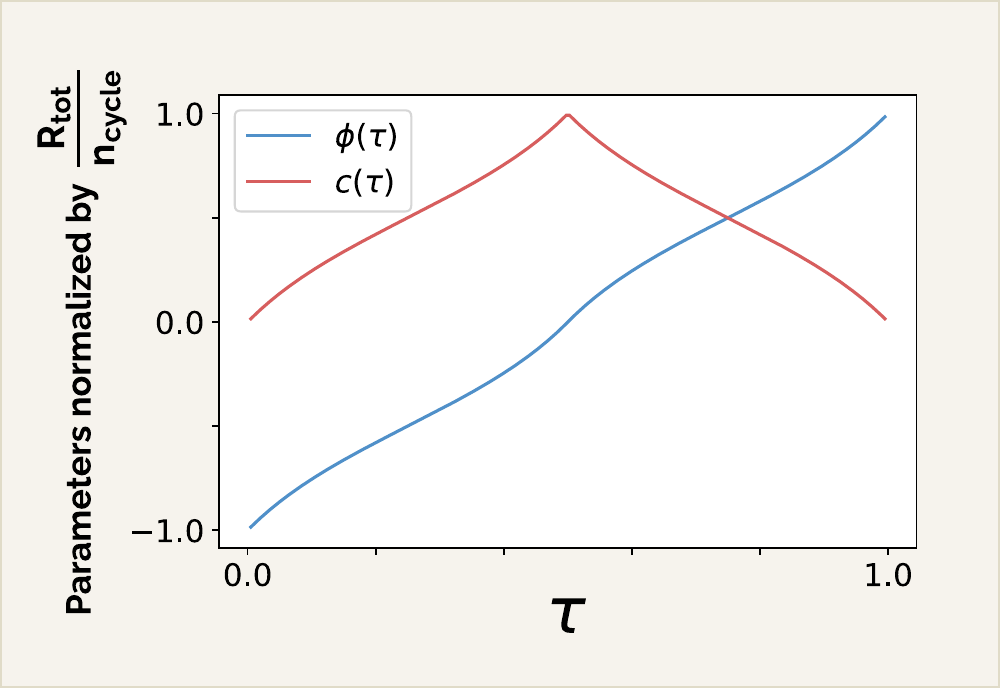}
    \caption{Values of $\phi$ and $c$ versus $\tau$ for the annealing protocol used in this study as defined in equations \ref{eq:phi_protocol} and \ref{eq:c_protocol}.}
    \label{fig:anneal_protocol}
\end{figure}

In principle, if performed slowly enough, any continuous ramp will successfully implement the transfer, however, with a finite number of cycles, the exact profile will matter. For the numerical experiments presented later in this work it is important to develop an annealing schedule which can be implemented sensibly for a wide range of cycle numbers and perform reasonably well. Fortunately we can take inspiration from the single avoided crossing model presented in other work \cite{Morley19a,Berwald2024Zeno} to develop an annealing profile which is optimal for a single mode. From the previous work in the literature, we know that such a protocol should obey 
\begin{equation}
    \frac{\phi_i}{c_i}=\cot\left(\pi\tau_i\right)
\end{equation}
where $0<\tau_i<1$ is a unitless time parameter which is linearly swept through the protocol. This still leaves one degree of freedom, the total amount of rotation performed at each step $|\phi|+|c|$, we elect to keep this constant throughout the protocol the protocol by choosing
\begin{align}
    \phi(\tau)=\frac{R_\mathrm{tot}}{n_\mathrm{cycle}}\frac{\cot\left(\pi\tau\right)}{1+|\cot\left(\pi\tau\right)|} \label{eq:phi_protocol} \\
    c(\tau)=\frac{R_\mathrm{tot}}{n_\mathrm{cycle}}\frac{1}{1+|\cot\left(\pi\tau\right)|} \label{eq:c_protocol}
\end{align}
where $R_{tot}$ is the total combined rotation angle and $n_\mathrm{cycle}$ is the total number of cycles. The time parameter\footnote{here we use the physicist's convention of one-based indexing $i\in\{1,2...n_\mathrm{cycle}\}$ as opposed to zero-based indexing commonly used in computer science.} is set to $\tau_i=\frac{i}{n_\mathrm{cycle}+1}$. This protocol is plotted in figure \ref{fig:anneal_protocol}.

\begin{figure}
    \centering
    \includegraphics[width=0.75\linewidth]{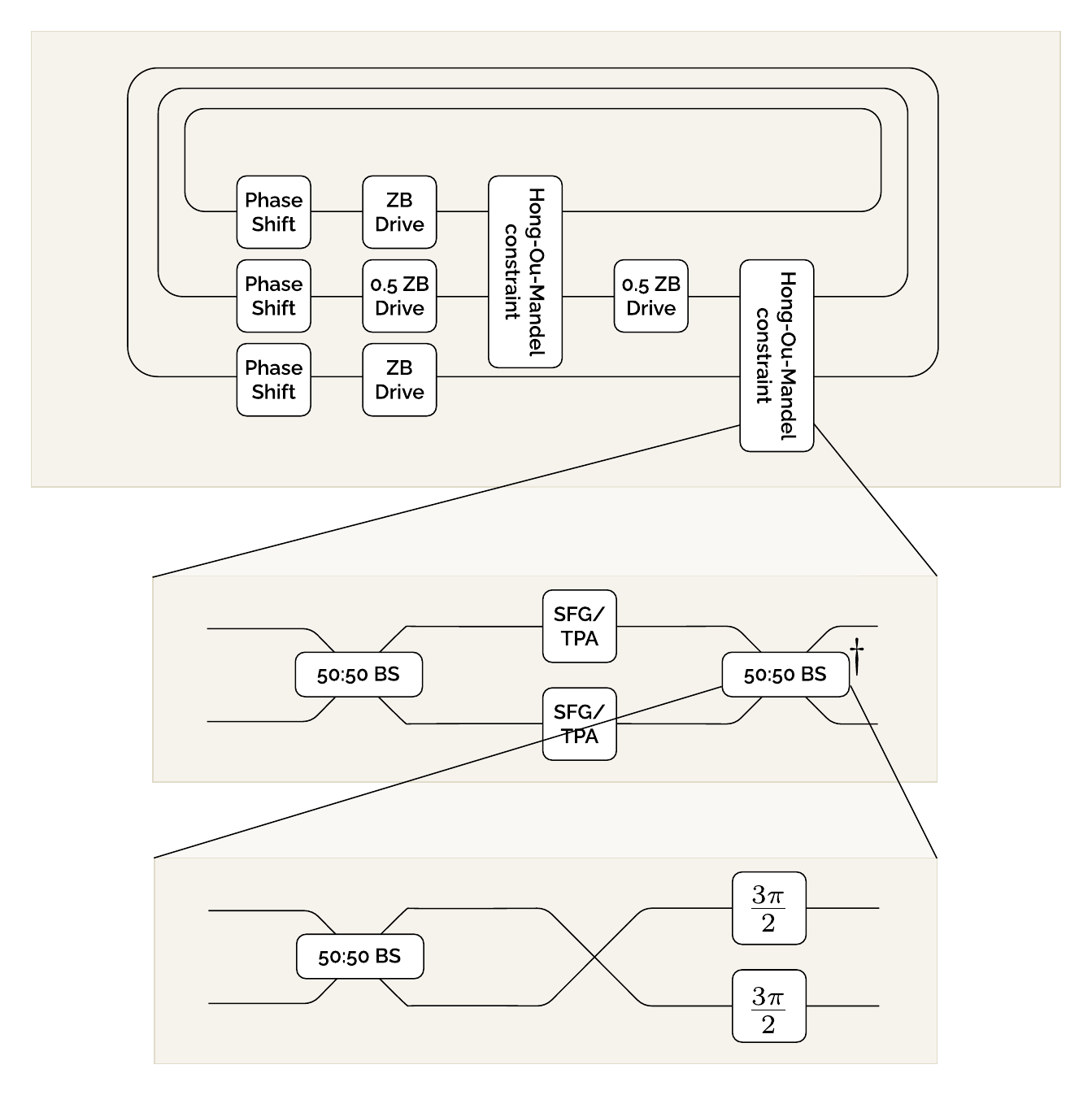}
    \caption{Block diagram of implementation of all-optical Zeno-effect-based annealing to solve the maximum independent set problem (specifically the three-node-line problem from equation \ref{eq:three_node_line}). In this case the phase shift per cycle and driving strength are updated slightly each cycle and the Hong-Ou-Mandel effect combined with sum-frequency-generation based phase shift enforce the constraint. Two-photon absorption could also be used in this configuration. ZB drive describes a Zeno-blockade based driving process as discussed in the main text.}
    \label{fig:block_diagram_coherent}
\end{figure}

By itself, this driving does not perform a useful computation, although it does in principle provide a recipe for a deterministic single-photon source. Since it is easier to understand, we first discuss conceptually how to implement a protocol where each of the modes used for encoding are separated spatially (see figure \ref{fig:block_diagram_coherent}). Such an implementation has the disadvantage of having a number of non-linear elements which scales as the number of edges in the graph, however as we discuss in section \ref{sub:time_bin_enc}, a time-domain encoding strategy can allow an equivalent protocol for an arbitrary size graph to be implemented with only two non-linear elements.

If an anneal is performed with constraints applied at each cycle, as depicted in figure \ref{fig:block_diagram_coherent}, than if evolved adiabatically (for example using the protocol in figure \ref{fig:anneal_protocol} with sufficient runtime), it will arrive at the state which achieves the lowest energy \emph{within the constrained subspace}. In this case, energy is determined by the number of modes in the $\ket{1}$ state. Maximizing this number subject to independence constraints is precisely the definition of the maximum independent set problem, which is NP-hard. We have therefore demonstrated an annealing technique where interactions are implemented through two-photon absorption. We discuss two natural generalisations to other problem classes, weighted maximum independent set and quadratic binary optimization in appendix \ref{app:gen_ext}.

 \subsection{Time-Domain Encoding \label{sub:time_bin_enc}}

While figure \ref{fig:block_diagram_coherent} depicts conceptually how to implement the protocol, it would involve constructing a large number of non-linear elements which would scale with the number of edges of the graph defining the independence constraints. A much more practical experimental implementation would be to instead encode each mode into a time bin rather than spatially separating them. Assuming we are able to perform fast switching between optical fibers, we are able to implement a protocol which is mathematically equivalent to figure \ref{fig:block_diagram_coherent}, but only requires two non-linear elements, one to implement either sum-frequency generation or two-photon absorption for the constraint, and one to perform driving using a Zeno blockade.

\begin{figure}
    \centering
    \includegraphics[width=0.75\linewidth]{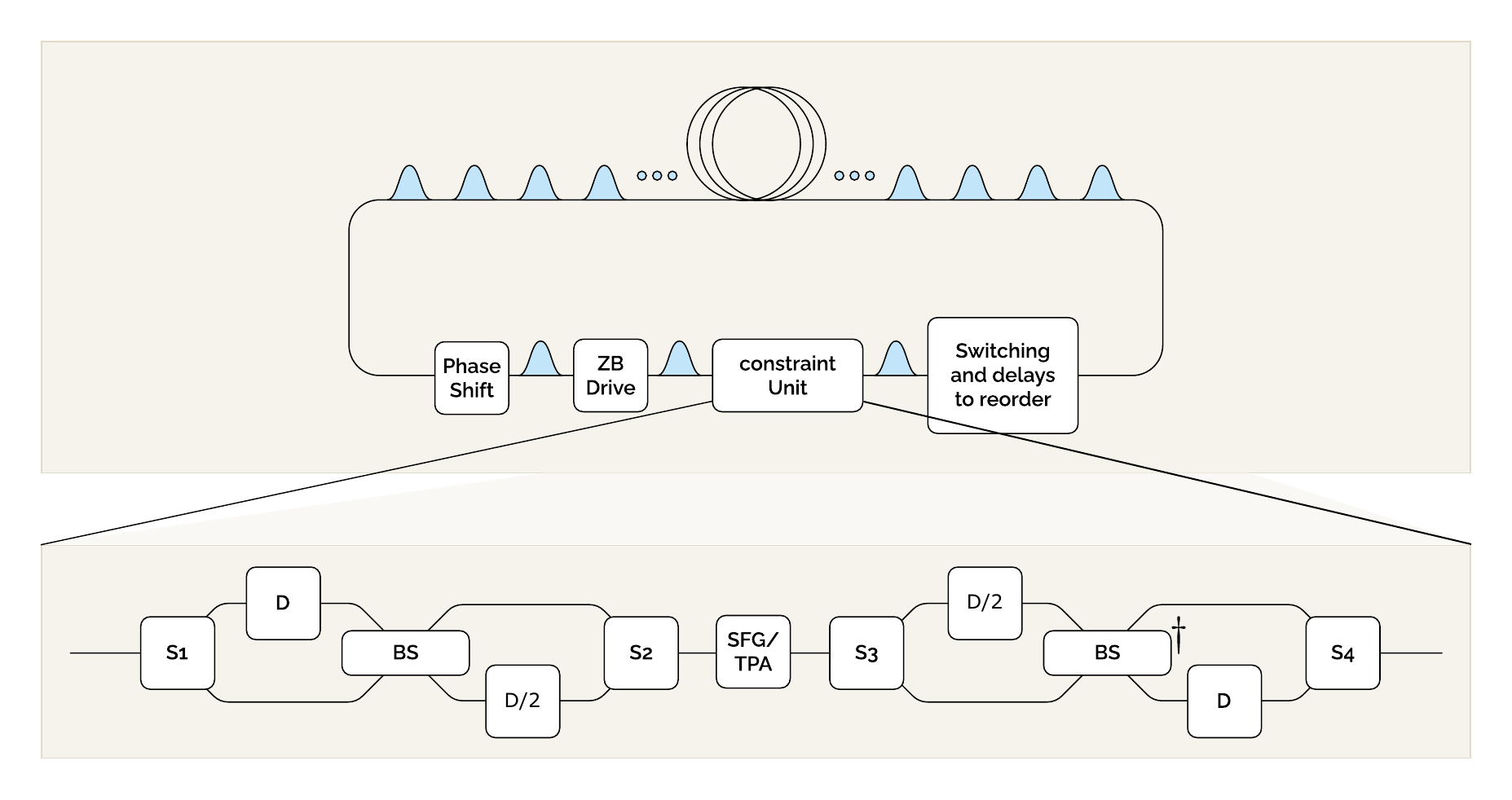}
    \caption{Implementation of the protocol shown in figure \ref{fig:block_diagram_coherent} using time-bin encoding. The lower diagram shows how to implement the interaction protocol using $50:50$ beamsplitters (BS), fast switches (S1,S2,S3,S4), and a non-linear element to implement SFG or TPA. Switching sequences to implement constraint and identity operations appear in table \ref{tab:sequence}.}
    \label{fig:time_bin_encoded}
\end{figure}

Figure \ref{fig:time_bin_encoded} shows a time-domain encoded implementation, in this implementation, each mode occupies a time bin and all time bins are separated by a delay $D$. Since the phase and driving operations are already independent on each mode, they can easily be moved to sequential operation in this setting. If fast switching is available, delay lines and switches can be used to reorder the time bins into arbitrary order after each pass. What is needed to achieve the interaction is then a unit which can be controlled to either interact a pair of modes separated by a delay $D$, or do nothing depending on the controls it is given. To allow sufficient space, a shorter delay of $\frac{D}{2}$ is needed for the inner application where the modes are sent through the non-linear element sequentially.

\begin{table}
\centering
\begin{tabular}{|c|c|c|c|c|}
\hline 
Bin & S1 & S2 & S3 & S4\tabularnewline
\hline 
\hline 
\multicolumn{5}{|c|}{Constraint}\tabularnewline
\hline 
First & 0 & 0 & 0 & 0\tabularnewline
\hline 
Second & 1 & 1 & 1 & 1\tabularnewline
\hline 
\multicolumn{5}{|c|}{Identity}\tabularnewline
\hline 
First & 0 & 0 & 0 & 0\tabularnewline
\hline 
Second & 0 & 1 & 1 & 0\tabularnewline
\hline 
\end{tabular}
\caption{Switching sequence for first and second time bins incident on each switch for the interaction implementation in figure \ref{fig:time_bin_encoded}. These implement either constraint or identity operations. In this sequence we take 0 to mean the top channel and 1 to mean the bottom channel. \label{tab:sequence}}
\end{table}

To implement an interaction, the first time bin needs to be delayed so that it is simultaneously incident on a $50:50$ beamsplitter with the time bin after, one output then needs to be delayed so that each can pass through the non-linear operation sequentially. For the second beamsplitting operation, the time bin which was not initially delayed needs to be delayed so they once more enter simultaneously. Finally, a delay must be applied to one time bin so that they are again sequential. This can be achieved using the optical circuit depicted in figure \ref{fig:time_bin_encoded} when switches S1-S4 all send the first time bin which is incident on them to the top channel, and the second to the bottom as shown in table \ref{tab:sequence}. In this operation, each time bin receives a total delay of $\frac{3}{2} D$ from it's initial position on the fiber. 

However, for an arbitrary graph, it is likely not the case that each adjacent pair of time bins should have an interaction between them. Fortunately a different control pattern can also implement an identity operation, which effectively only delays each time bin by $\frac{3}{2} D$. To achieve this, the first switch, S1, can just always send the time bin to the top channel, where it will be delayed, if the next time bin in the sequence is also in the top channel, than vacuum will be incident on the lower channel. S2 can then be used to send each time bin produced by the beamsplitter through the non-linear operation. However, since there is only a maximum of one photon between the two time bins, the non-linear element acts as the identity. The other switch S3, will then ensure that the two time bins are again simultaneously incident on the second beamsplitting operation. Since the two operations are the inverse of each other by definition, this results in the top mode with the original contents of the time bin and vacuum being sent to the (unused) delay line. To complete the protocol S4 is set to receive the top mode both times. This sequence is also shown in table \ref{tab:sequence}

Given an order of the time bins, whether an interaction is initiated can be determined from the edges of the graph, if there is an interaction between the node in time bin $t_j$ and the previous time bin $t_{j-1}$, than $t_j$ should be sent to the bottom (undelayed) channel, initiating an interaction, and S4 should be adjusted appropriately so that both bins are captured. For a highly (or fully) connected graph, delays should be performed so that all nodes are adjacent to each other at least once in a cycle, an example of a protocol which accomplishes this can be found in appendix \ref{app:delay_reorder}. For sparser graphs, what reordering to apply is a non-trivial compilation problem, and one which is beyond the scope of the current work.

\section{Performance Results}

\subsection{Single-mode driving\label{sub:single_mode_drive}}

We first turn our attention to the Zeno effect required to confine the system in the $\{\ket{0},\ket{1}\}$ subspace while coherently driving the photon number. Figures \ref{fig:Zeno_onset_TPA} and \ref{fig:Zeno_onset_SFG} suggest that a coherent Zeno effect my be more effective for this purpose than an incoherent one. To systematically study this difference, and to understand the behavior when a Zeno effect is partially coherent, we study the action of sum-frequency generation with a lossy pump mode as described in equation \ref{eq:Omega_dl_SFG}. This allows us to interpolate between completely coherent Zeno effects $\eta=0$, and completely incoherent, we which sets in when $\eta=4 \sqrt{2} \gamma$ as we demonstrate in appendix \ref{app:lossy_SFG_coherence}. Intermediate values correspond to partial coherent, where some amplitude will return to $\ket{2}$ and higher modes from the pump mode, but not completely. 

To systematically study the ability of a Zeno effect to contain the system in $\{\ket{0},\ket{1}\}$, we consider applying the superoperator in equation \ref{eq:Omega_dl_SFG} to the vacuum for a time of $t=\frac{\pi}{2c}$. This time is much longer than we would apply a single round of driving for in the real protocol, but provides a convenient way to measure performance. For a perfect Zeno blockade the system will end completely in the $\ket{1}$ state after this operation, whereas for no blockade, the probability will only be $\approx 0.2$ (see figure \ref{fig:Zeno_onset_TPA}). 

\begin{figure}
    \centering
    \includegraphics[width=0.5\linewidth]{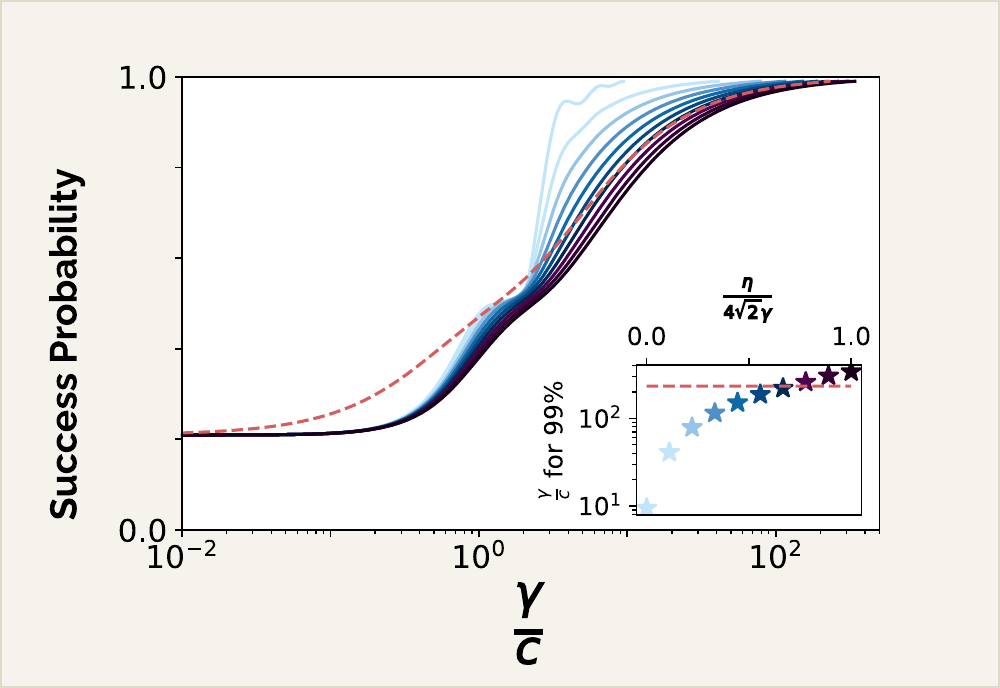}
    \caption{Probability of being in the $\ket{1}$ state after applying $\Omega_\mathrm{dl,SFG}(c,\gamma,\eta,t=\frac{\pi}{2c})\left[\ketbra{0}{0}\right]$ from equation \ref{eq:Omega_dl_SFG}. We vary values of $\frac{\eta}{\gamma}$ between undamped ($\eta=0$) and critically damped ($\eta=4\sqrt{2}\gamma$). Note that curves end when the $99\%$ threshold is reached for increased visibility of the approach of the other curves and to save compute time. The inset shows the value of $\gamma$ to reach $99 \%$ success probability versus $\frac{\eta}{4 \sqrt{2}\gamma}$ (using the same color scheme). The red dashed line shows the result for two-photon absorption with a loss rate of $\gamma$ for comparison. The main Hilbert space is truncated at $10$ photons for $\gamma<10$ and $5$ photons for greater values. The pump Hilbert space is limited to the maximum number of pump photons which could be generated.}
    \label{fig:incohere_cohere_drive}
\end{figure}

Figure  \ref{fig:incohere_cohere_drive} shows the result. Firstly we notice that a completely coherent Zeno effect can reach $99\%$ success probability with a $\gamma$ value which is more than an order of magnitude smaller than in the completely incoherent case. However we can also make an interesting observation that partially coherent operation can still have a significant advantage over incoherent operation. This indicates that even partially retaining the pump mode is a desirable goal at least in the range between perfect coherence and critical damping. Finally, we notice that while two-photon absorption behaves qualitatively differently even than lossy sum-frequency generation at the critical damping value, the time to reach $99\%$ success probability is not substantially different\footnote{At least when compared to other factors, like coherence levels of the Zeno effect}. This difference is caused by the fact that two-photon absorption is a Markovian model, while even at the critical damping point, lossy sum-frequency generation acts in a highly non-Markovian way. 

\begin{figure}
    \centering
    \includegraphics[width=0.5\linewidth]{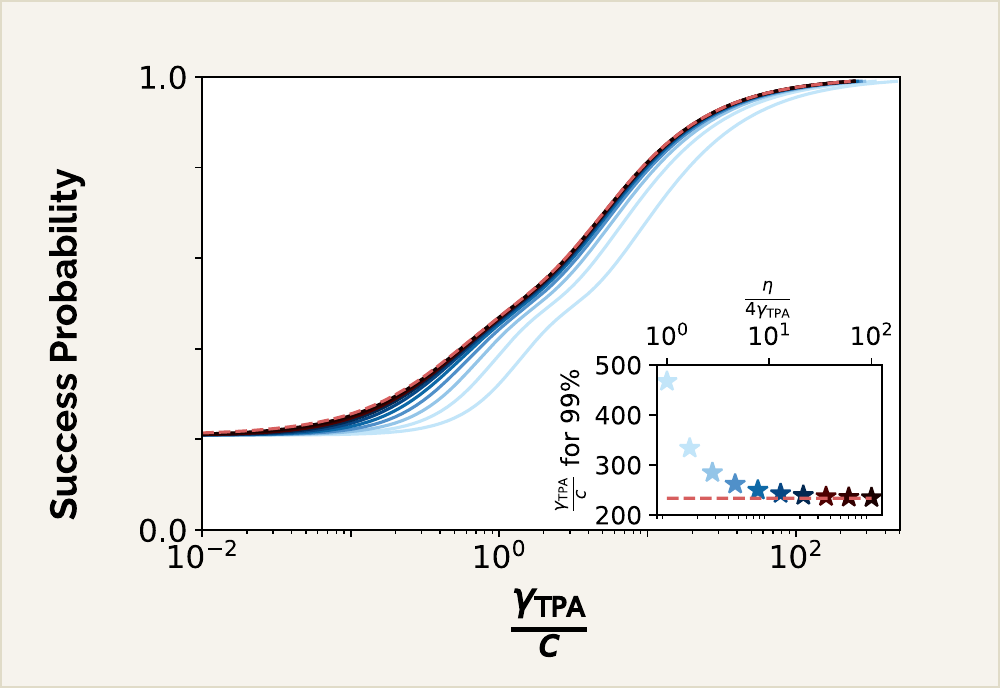}
    \caption{Probability of being in the $\ket{1}$ state after applying $\Omega_\mathrm{dl,SFG}(c,\gamma,\eta,t=\frac{\pi}{2c})\left[\ketbra{0}{0}\right]$ from equation \ref{eq:Omega_dl_SFG}. We vary values of $\frac{\eta}{\gamma}$ between critically damped ($\eta=4 \gamma_\mathrm{TPA}$) and strongly over damped ($\eta=400 \gamma_\mathrm{TPA}$). We vary $\gamma_\mathrm{TPA}$, the effective decay rate for two-photon absorption rather than $\gamma$ to allow a direct comparison of the curves, the underlying mathematics can be found in appendix \ref{app:lossy_SFG_coherence}, and the specific formula for $\gamma$ is equation \ref{eq:gamma_TPA}. Note that curves end when the $99\%$ success threshold is reached for increased visibility of the approach of the other curves and save compute time. The inset shows the value of $\gamma_\mathrm{TPA}$ to reach $99 \%$ success probability versus $\frac{\eta}{4 \gamma_\mathrm{TPA}}$  (using the same color scheme). The red dashed line shows the result for two-photon absorption with a loss rate of $\gamma_\mathrm{TPA}$ for comparison. The main Hilbert space is truncated at $10$ photons for $\gamma<10$ and $5$ photons for greater values. The pump Hilbert space is limited to the maximum number of pump photons which could be generated.}
    \label{fig:markov_appraoch_drive}
\end{figure}

 Markovian two-photon absorption is slightly more effective than the critically damped sum-frequency generation at implementing a Zeno blockade. This can be understood intuitively because far from the Markovian limit the pump mode must become occupied before amplitude can be removed from the $\ket{2}$ state, whereas maximum absorption happens immediately in the Markovian limit. Intuitively, since loss from the $\ket{2}$ state enacts the Zeno effect, an initial slowing down slightly reduces the effectiveness. Note however that this effect is significantly weaker than the effect of partial coherence on the strength of the Zeno blockade. Figure \ref{fig:markov_appraoch_drive} shows the effect of approaching the Markovian limit with an effectively constant two-photon absorption rate.\footnote{It is worth emphasising that in figure \ref{fig:markov_appraoch_drive}, $\gamma$ is effectively scaled with $\sqrt{\eta}$ so that the approached loss rate is finite, just increasing $\eta$ with the coupling unchanged would result in $\gamma_\mathrm{TPA}\rightarrow 0$, see appendix \ref{app:lossy_SFG_coherence} for details.}

\subsection{Coherent versus Incoherent Constraints}

For the constraints implemented as depicted in figure \ref{fig:block_diagram_coherent} a separate process active at the same time as sum-frequency generation or two-photon absorption, the exact details of how the process occurs do not matter, only the level of coherence retained at the end of the process, for this reason, we can consider the effect of coherent versus incoherent constraints entirely by examining the operation of ideal sum-frequency generation with an initially empty pump mode which is traced out at the end of the evolution described by equation \ref{eq:G_sfg}. 

To study an interpolation between fully incoherent and fully incoherent operation, we can examine the range of $\frac{\pi}{4 \sqrt{2}} \le\gamma t_\gamma\le \frac{\pi}{2 \sqrt{2}}$. For $\gamma t=\frac{\pi}{4 \sqrt{2}}$, the $\ketbra{2,0_p}{2,0_p}$ state will fully be converted to $\ketbra{0,1_p}{0,1_p}$. Likewise $\ketbra{2,0_p}{1,0_p}$ will be converted to $i\ketbra{0,1_p}{1,0_p}$. After tracing over the pump modes results in $\ketbra{2}{2}\rightarrow \ketbra{0}{0}$ and $\ketbra{2,0_p}{1,0_p}\rightarrow 0$, the same as applying two-photon absorption for a long period of time. On the other hand for $\gamma t=\frac{\pi}{2 \sqrt{2}}$, the pump mode photon will fully return $\ketbra{2,0_p}{2,0_p}\rightarrow \ketbra{0,1_p}{0,1_p} \rightarrow \ketbra{2,0_p}{2,0_p}$, however we also observe $\ketbra{2}{1}\rightarrow \ketbra{2,0_p}{1,0_p}\rightarrow i\ketbra{0,1_p}{1,0_p} \rightarrow -\ketbra{2,0_p}{1,0_p} \rightarrow -\ketbra{2}{1}$, leading to an effective phase inversion. Intermediate values will lead to a mixture of the loss and phase inversion behavior, so we can capture the full range of coherent and incoherent behavior by examining this range of values. Full phase inversion is not always desirable, so it additional phase shifting may be applied to the pump mode, this however does not have an effect on the interpolation between coherent and incoherent behavior.

\begin{figure}
    \centering
    \includegraphics[width=0.5\linewidth]{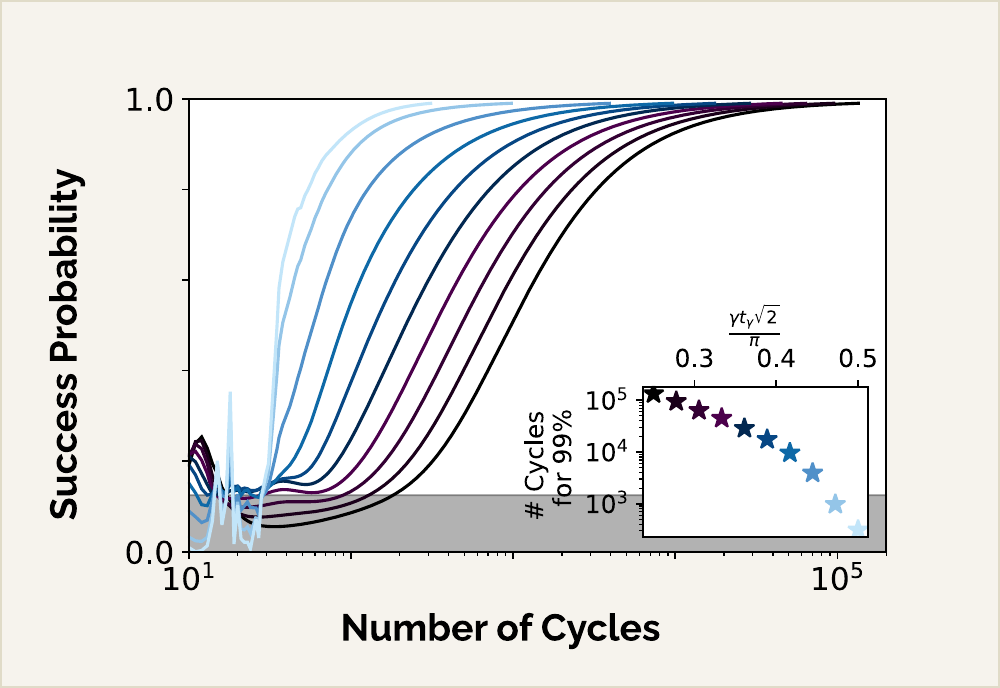}
    \caption{Probability of finding the maximum independent set for the three node line graph described in equation \ref{eq:three_node_line} using the annealing procedure described in figure \ref{fig:block_diagram_coherent} and the annealing schedule depicted in figure \ref{fig:anneal_protocol} with a total rotation of $R_\mathrm{tot}=20\pi$. For each gadget we applied equation \ref{eq:master_constraint} with $\phi_Q=0$ and $\eta t=0$. The number of cycles is varied until $99\%$ success is achieved, and a range of $\frac{\pi}{4 \sqrt{2}} \le\gamma t\le \frac{\pi}{2 \sqrt{2}}$ are used for sum-frequency generation when implementing the constraint. Driving is assumed to operate ideally in $\{\ket{0},\ket{1}\}$. The gray shaded region is the region where the success probability is less than the $\frac{1}{8}$ likelihood attained from random guessing. The inset shows the number of cycles needed to attain $99\%$ success versus $\gamma t$ and uses the same color scheme as the main figure. Note the logarithmic scale on the x-axis of the main figure.}
    \label{fig:interaction_coherence_scaling}
\end{figure}

\begin{figure}
    \centering
    \includegraphics[width=0.5\linewidth]{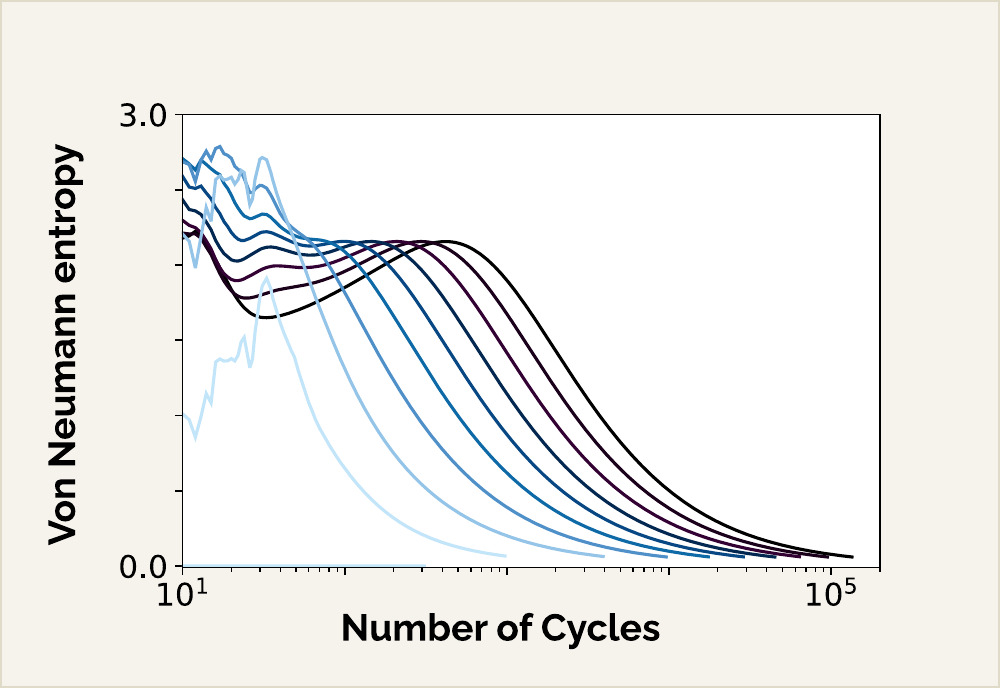}
    \caption{Coherence measured by von Neumann entropy (note 0 is coherent and higher value are less coherent) for the same protocols as figure \ref{fig:interaction_coherence_scaling}. These protocols solve the maximum independent set problem in \ref{eq:three_node_line} using the annealing procedure described in figure \ref{fig:block_diagram_coherent} and the annealing schedule depicted in figure \ref{fig:anneal_protocol} with a total rotation of $R_\mathrm{tot}=20\pi$. For each gadget we took $\phi_Q=0$, such that it yields a zero phase. The number of cycles is varied until $99\%$ success is achieved, and a range of $\frac{\pi}{4 \sqrt{2}} \le\gamma t\le \frac{\pi}{2 \sqrt{2}}$ are used for sum-frequency generation when implementing the constraint. Driving is assumed to operate ideally in $\{\ket{0}\ket{1}\}$. The colors used match figure \ref{fig:interaction_coherence_scaling}. Note that for fully coherent constraints the entropy is identically zero independent of the number of cycles. Note the logarithmic scale of the x-axis on the main figure.}
    \label{fig:interaction_coherence_entropy}
\end{figure}

Figure \ref{fig:interaction_coherence_scaling} shows the results, which parallel the results for driving in figure \ref{fig:incohere_cohere_drive}; a coherent Zeno effect leads to drastically better performance, and even partially coherent operation is beneficial. When an incoherent Zeno effect is not completely effective, the system experiences decoherence, this can be quantified by computing the von Neumann entropy of the system
\begin{equation}
s_\mathrm{VN}=-\mathrm{Tr}\left[\hat{\rho}\log_2\left(\hat{\rho} \right)\right]
\end{equation}
where a von Neumann entropy of $0$ corresponds to a pure state, an a maximally-mixed three-qubit state will have $s_\mathrm{VN}=3$. An advantage of using a base-two logarithm is that a maximally mixed state with $n$ qubits will have  $s_\mathrm{VN}=n$. Figure \ref{fig:interaction_coherence_entropy} shows the evolution of the von Neumann entropy for the corresponding protocols in figure \ref{fig:interaction_coherence_scaling}, and illustrates that an incomplete partially-coherent Zeno effect can lead to substantial decoherence.

\subsection{Approach to Ideal Constraints}

\begin{figure}
    \centering
    \includegraphics[width=0.75\linewidth]{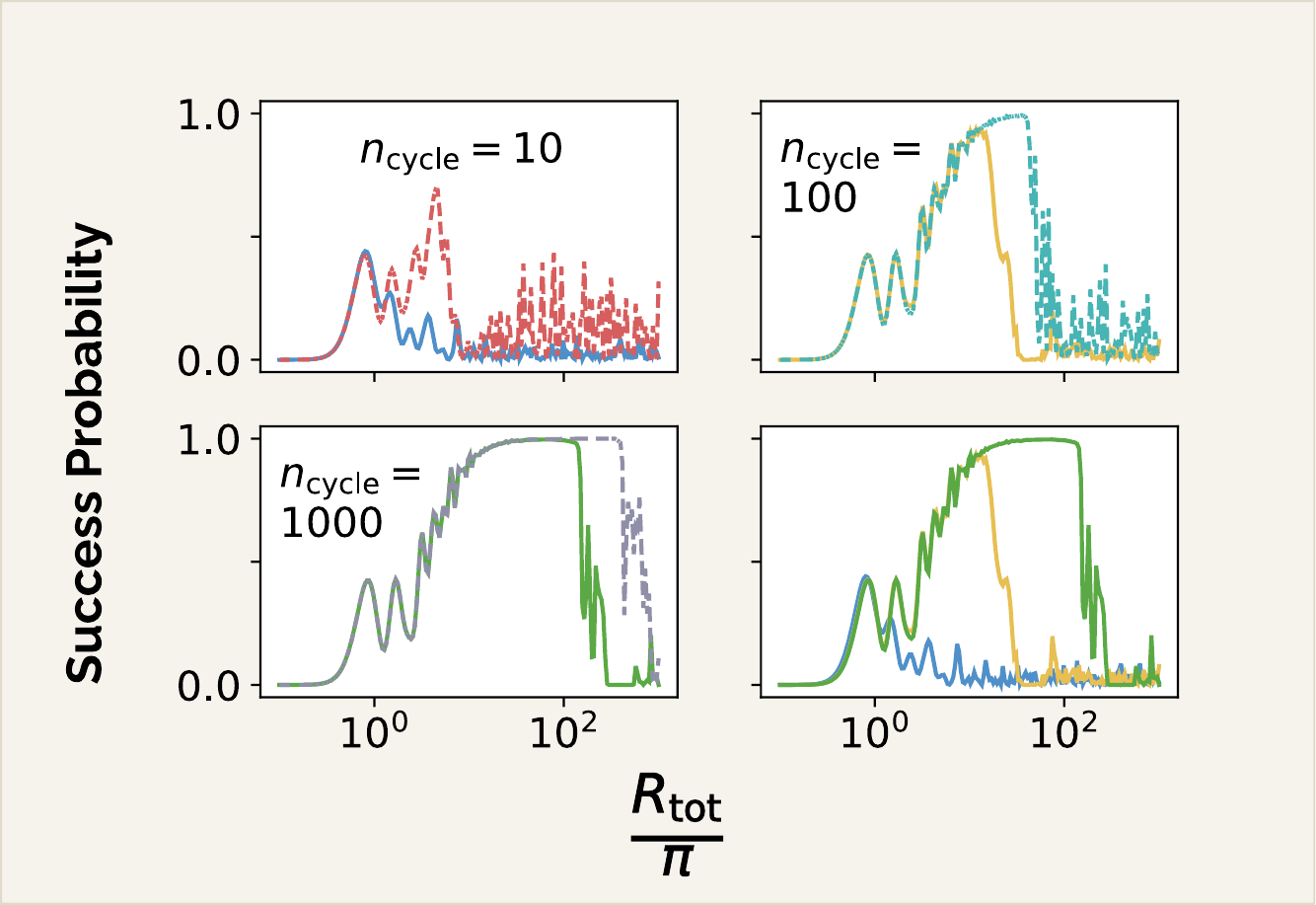}
    \caption{Success probability versus $R_\mathrm{tot}$ ideal (dashed lines) versus phase based (solid lines) Zeno constraints for finding the maximum independent set of the $5$ node graph given in equation \ref{eq:five_node_graph}. These results were obtained using a state-vector simulation. In the ideal case the drivers were modified to not allow transitions to non-independent sets, in the phase based case the protocol described in \ref{fig:block_diagram_coherent} was performed, with $\phi_Q=\frac{3\pi}{2}$, for a total phase kick of $\frac{\pi}{2}$. The lower-right figure compares phased-based effects, and uses the same color scheme as the other figures for $n_\mathrm{cycle}$. Note Logarithm scale on x-axes. }
    \label{fig:ideal_vs_Zeno_5QB}
\end{figure}

A strong enough Zeno effect is able to effectively confine the system within the subspace occupied by independent sets. Given that the strength of the constraint terms does not degrease as $n_\mathrm{cycle}$ is increased, while the rotation and phase in each cycle does, for enough cycles the dynamics should be perfectly confined. This is in contrast to transverse field Ising annealing, where quadratic penalties enforce the constraints. This can be seen in figure \ref{fig:ideal_vs_Zeno_5QB}, where a coherent Zeno effect can match perfectly implemented constraints which are simulated by removing transitions to non-independent sets. 

From figure \ref{fig:ideal_vs_Zeno_5QB} we see that until a critical value of $R_\mathrm{tot}$ the behavior of a phase-based Zeno effect matches perfectly the ideal constraints. This value depends linearly on $n_\mathrm{cycle}$ as the amount of rotation in each cycle is inversely proportional to $n_\mathrm{cycle}$, and when the rotation in a cycle becomes too large, it overwhelms the phase effect of the constraints, at which point the algorithm effectively fails.

\section{Numerical Methods}

All numerical calculations were performed using Python. We particularly made use of the the numpy \cite{numpy}, scipy \cite{2020SciPy-NMeth}, and matplotlib packages \cite{hunter2007matplotlib} as well as Jupyter notebooks for plotting \cite{jupyter}. The code used for this paper and generated data are publicly available at \url{https://github.com/qci-github/eqc-studies/tree/main/optical_annealing_code} and \url{https://doi.org/10.6084/m9.figshare.31362496}. All simulations were performed on a consumer model laptop computer, and no HPC resources were used, but some computations took over 24 hours to complete. Several pre-computation strategies were employed to get the simulations to run in a reasonable timescale, these are discussed in appendix \ref{app:detailed_num}.

\section{Discussion}

In the previous sections we have laid out a way to implement an optical optimization protocol to solve a maximum independent set. We show that this can be done both in a coherent setting, using entirely incoherent two-photon absorption mechanisms, as well as in intermediate regimes where partially coherent effects act. We generally have found numerically that more coherent operation leads to better performance, but even a fully incoherent model does work. This is interesting because in the fully incoherent limit, the only nonlinearity in the model is coming from loss mechanisms, which are conventionally viewed as undesirable. This in contrast to other models of computing like the KLM protocol \cite{Knill2001KLM} where measurement and feedback are required. On the other extreme when fully coherent, the protocol we describe here can be described as an implementation of a gate-model algorithm, with the interactions being non-linear optics implementations of controlled phase gates. In the fully-coherent extreme, the architecture discussed here in principle has a universal gate set, where arbitrary $X$ rotations can be achieved by multiple passes through the Zeno blockade driver $Z$ rotations through phase shifting, and entanglement through the constraint unit.

Furthermore, while the results we have presented have all focused on annealing-like protocols similar to the gate model AQA algorithm proposed in \cite{Willsch2022AQA} with fixed phasing and driving schedules, the protocol proposed here could be applied in a variational way, as is done in the quantum approximate optimization algorithm (QAOA)\cite{Farhi14a,Hadfield2019QAOA}. For an extensive discussion of the benefits and drawbacks of AQA, see \cite{Willsch2022AQA}. It would also be interesting to see if the protocol proposed here could be modified so that the rotation and phasing are simulataneous, giving a protocol more like a multi-stage quantum walk \cite{Callison21a,Banks2024continuoustime,Hopkins2025MSQW,Hattori2025MSQW}, which have in practice shown benefits of Trotterised QAOA-like evolution \cite{Gerblich2024MSQWvsQAOA}.

The possibility of a protocol with partial coherence however is interesting because it suggests that sum-frequency generation where the pump mode is imperfectly retained can still be useful, and indeed more useful than one where it is immediately lost. Since for example it is hard to implement a cavity system which is resonant with both the pump and cavity modes \cite{Chen2018NanoCircuits,Chen2021PhotonConversion,Chen2021FrequencyDouble}, not restricting that the pump mode is perfectly retained (as it would be for a controlled-phase gate implementation) is likely to make the protocol easier to implement in practice.  There may be interesting connections with efforts to use Zeno effects (or related weak measurement techniques \cite{Stollenwerk2025Measurement}) for constraints in a gate model context \cite{Benjamin2017MeasurementSAT,Herman2023constrainZeno,Zhang2024MeasrurementSAT} however.

When the protocol proposed here is implemented correctly, the system remains in a constrained subspace throughout the entire protocol. Moreover, the phases used in conjuction with driving only act linearly on individual qubits, rather than quadratically. This is in stark contrast with traditional transverse-field Ising annealing (or QAOA based on digitized versions of it), where the system initially starts in a superposition of all computational basis states and the constraints must be enforced by strong quadratic penalties. There has been work to understand subspace preserving drivers in other contexts \cite{Hen2016Constrained,Hadfield2019QAOA,chancellor2019,Wang2020XYQAOA,liu2024comparisonconstrainencodingmethods}, but here we present a paradigm where the constraints are a fundamental part of the paradigm\footnote{It is worth remarking that Rydberg based annealing treats constraints similarly to how we do here, with them being a fundamental part of the paradigm.}.

While the protocol discussed here is very different from transverse-Ising annealing, the implementation with linear terms and constraints is exactly the way in which linear programming problems are stated in classical computing and operations research. Conceptually, the solving here is similar to an interior-point method \cite{Rardin1998OpResearch,Wright2004Interior}, where the phases which are applied act as an effective barrier function and the quantum states are forced against a constraint in superposition. The key difference here is that interior-point methods are typically applied to continuous problems, rather than the discrete domain considered here. Given the relative maturity of classical integer linear programming algorithms \cite{Rardin1998OpResearch} compared to quadratic solvers, the statements of problems in terms of linear penalties and constraints, rather than quadratic interactions likely provides an interesting route for hybrid quantum/classical algorithms. 

\subsection{Toward Error Mitigation}

To establish the theoretical underpinnings of the optical protocol we have proposed here we have worked in an idealized setting, in particular we have ignored the potential effects of single-photon loss in the qubit modes and partial distinguishability between modes, which would lead to an incomplete Hong-Ou-Mandel effect. Errors from modes being partially distinguishable would cause the constraints to probabilistically fail during a particular round of the protocol. However, given the nature of the Zeno effect, this would only lead to a small increase in amplitude outside of the constrained subspace, which would lead to a low probability of failure of the protocol overall. A full analysis is beyond the scope of the current work, but conceptually small imperfections which can probabilistically cause distinguishability (timing errors for example) are unlikely to totally ruin the Zeno effect.

Qubit photon loss on the other hand effectively leads to removing a node from the independent set, and is a serious logical error. Beyond a certain scale, some kind of error mitigation strategy will become necessary. A strategy which has been well explored in the quantum annealing setting is to consider multiple interacting copies of of a problem Hamiltonian such that they either reinforce the correct answer, or so that frustration helps to contain errors \cite{Pudenz2014AnnealCorrect,Vinci2015AnnealCorrectMinor,Bennett2023QWCorrect,Bennett2025CorrectLHZ,Hattori2025AnnealingCorrect}. In fact error mitigation strategies was shown to improve the level of approximation at which an advantage can be seen \cite{Munoz-Bauza2025AnnealingAdvantage}. A major problem which traditional quantum annealing has which would not be present here is the issue of combining correction strategies with problem mapping strategies such as minor embedding, \cite{Vinci2015AnnealCorrectMinor,Bennett2025CorrectLHZ}.

One advantage of mitigating against photon-loss error in this setting is that this error only occurs one way, it only leads to a smaller set, never a larger one. For this reason an effective strategy may be to take multiple copies of the same problem, where each edge effectively becomes a fully connected bipartite graph between the nodes in each copy, mathematically, if the original graph is defined by a set of edges $E$, and the vertices are two-index objects $V_{j,k}$, where $j$ is the index within the original graph, and $k$ is the index of the copy (with $n_\mathrm{copy}$ total copies), the edges of the new graph would become
\begin{equation}
    E^{(enc)}=\bigcup_{j,k\in E}\bigcup^{n_\mathrm{copy}-1}_{p=0}\bigcup^{n_\mathrm{copy}-1}_{q=0}\{\{V_{j,p},V_{k,q}\}\}, \label{eq:Err_mit}
\end{equation}
where $\bigcup$ indicates a repeated union operation between the set of edges. To visualize this we have shown an encoding of three copies of the three node-line graph in figure \ref{fig:MIS_graphs}. 

\begin{figure}
    \centering
    \includegraphics[width=0.5\linewidth]{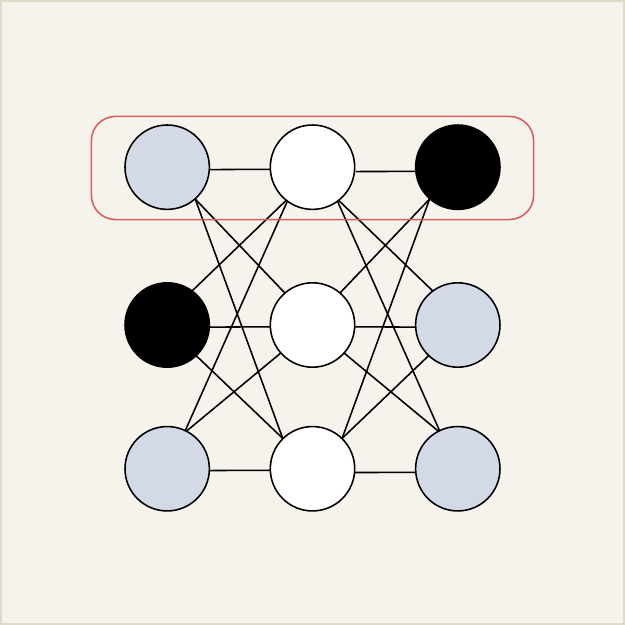}
    \caption{Independent set graph resulting from application of the error mitigation strategy from equation \ref{eq:Err_mit} applied to finding the maximum independent set of the three-node-line graph defined in equation \ref{eq:three_node_line} with $n_\mathrm{copy}=3$. One of the copies is circled and the grayed out nodes indicates a pattern of photon loss where the correct independent set would be recoverable despite $\frac{2}{3}$ of the photons being lost.}
    \label{fig:Error_mitigate}
\end{figure}

If we only consider photon loss as a source of error, then even a single photon in one copy indicates that a node should be included in the independent set. Moreover, the independence constraint for all copies will be maintained in this case. This fact suggests that the best decoding strategy is to treat a vertex as part of the set if that vertex is in the set in one or more copies. A full analysis of this strategy is beyond the scope of the current paper, but these facts suggest that it may be effective. It would be interesting to explore if there is a way to deterministically recover from the photon loss and therefore the loss will still introduce entropy into the system, which will lead to decoherence between the logically valid states. 

\section{Conclusion and Future Directions}

We have shown an optical protocol which implements real-time evolution within the entropy computing paradigm. This can be equivalently considered as an optical method to implement quantum annealing. This work provides an exciting starting point to further develop the entropy-computing paradigm from a theoretical direction.

In the direction of implementation, it would be interesting to relax some of the requirements, for example the requirement for fast switching in the time-domain encoding. Likewise, it would be interesting to understand how to bridge between the encoding here and the differential encoding used in current entropy computers and discussed theoretically in \cite{nguyen2025entropycomputing,chukwu2025opticalQuantum}. In general, bringing the theory presented here and the physical implementations closer to each other provide a promising route for develop the entropy-computing paradigm. Likewise, it would be interesting to understand whether other theoretical implementations of fully-quantum entropy computing exist. Of particular interest would be to more concretely analyze the construction of an entropy computer based on imaginary, rather than real time evolution as discussed in the supplemental material of \cite{nguyen2025entropycomputing}.

Another direction for future work is understanding error mitigation, while we have discussed the potential for a scheme to mitigate against photon loss, which we expect to be the dominant error type, we have not fully analyzed it. Likewise a more extensive understanding of other error sources, how severe they are, and how they can be mitigated, would be useful for fully developing the paradigm.

Furthermore, independent set constraints can be viewed as a subset of linear constraints, notably a subset which involve exactly two variables. In principle NP-hardness of the problem implies all optimization problems could be mapped by reductions. However, it would be interesting to understand if the techniques proposed here could be generalized to other forms of linear constraints.

\section{Acknowledgments}

The authors thank Yu-Ping Huang and Andrew P. Rotunno for useful discussions. The authors also thank Joel Russell Huffman for improving the visual aspects of the figures and Po-Jen Wang for independently verifying some of the simulation results. All authors were fully supported by QCi in performing this work, in particular NC received no UKRI support in writing this paper.

\appendix

\section{Generalizations and Extensions\label{app:gen_ext}}

There are two natural extensions to the protocol given in sections \ref{sec:subspace_confine} and \ref{sec:opt_protocol}, for different kinds of problems. The first is a minor modification to the driving which allows the members of the independent set to be weighted as described in \ref{sub:wmis}. The second, is to convert the constraints to interactions and implement a quadratic binary model, it is explained in \ref{sub:quadratic}.

\subsection{Variable phasing for Weighted Maximum Independent Set \label{sub:wmis}}

\begin{figure}
    \centering
    \includegraphics[width=0.5\linewidth]{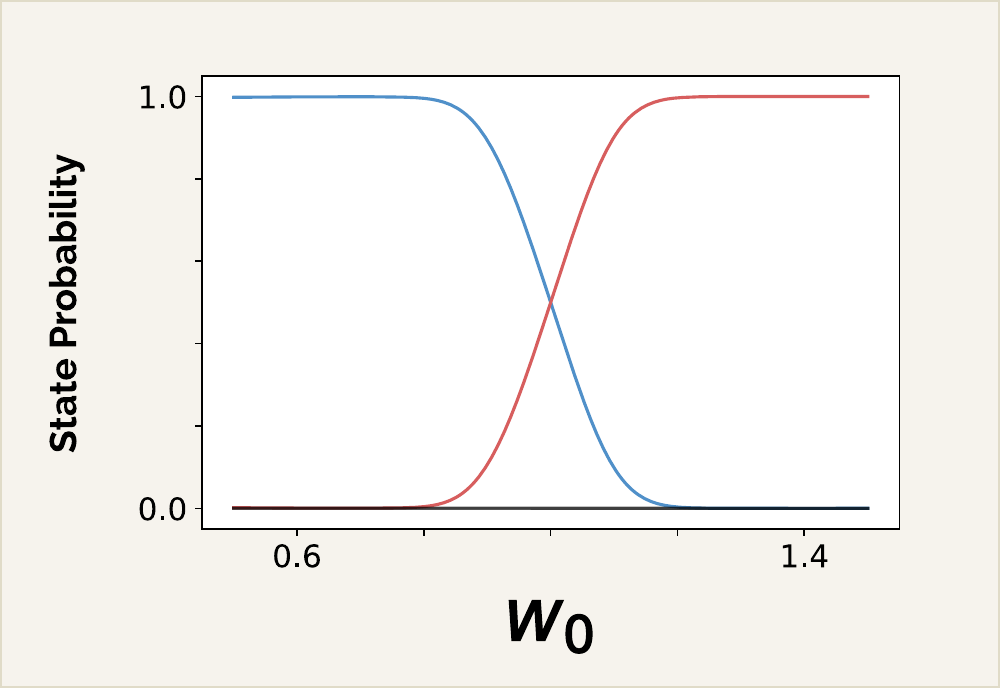}
    \caption{State probability found after a coherent anneal using the setup depicted in figure \ref{fig:block_diagram_coherent} with $n_\mathrm{cycle}=1000$, $R_\mathrm{tot}=200\pi$ and following the protocol defined in equations \ref{eq:phi_protocol} and \ref{eq:c_protocol}} with weights defined in equation \ref{eq:wmis_protocol}. The weight $w_0$ is varied while $w_1=1$ is held fixed. The blue line is the probability to be in the $\ket{01}$ state, while the yellow line is the probability to be in the $\ket{10}$ state, the probability for the other two states are shown as semi-transparent black lines. A perfect Zeno blockade is assumed in the driving for this protocol.
    \label{fig:wmis_swap}
\end{figure}

We now consider what happens if we apply variable amounts of phase to different modes in the encoding, by defining 
\begin{equation}
    \phi_i(\tau)=w_i\phi(\tau) \label{eq:wmis_protocol}
\end{equation}
where $\phi(\tau)$ is the schedule defined in equation \ref{eq:phi_protocol}. Assuming $w_i>0 \forall i$, then the vacuum will still be the initial ground state, but the excited state will be the one which maximizes $\sum_iw_in_i$, as in the definition of a weighted maximum independent set given in equation \ref{eq:wmis_def}, where is the photon number in mode $i$. As an example, we consider the weighted maximum independent set on a two-node graph $V=\{0,1\}$, $E=\{\{0,1\}\}$ where $w_0$ is varied and $w_1=1$ is held fixed, the results can be seen in figure \ref{fig:wmis_swap}.

\subsection{Extension to quadratic binary annealing \label{sub:quadratic}}

A final extension which we will discuss is how the Hong-Ou-Mandel effect based gadgets in figure \ref{fig:block_diagram_coherent} could be extended to encode general quadratic unconstrained binary (QUBO) problems. A QUBO is and unconstrained problem where the goal is to minimise 
\begin{equation}
    E_Q=\sum_j \sum_k s_js_kQ_{jk}
\end{equation}
where $Q$ is the matrix which defined the problem and as with the maximum independent set problem $s_j$ are binary variables. While the NP-hardness of (weighted) maximum independent set implies that a QUBO can be formally reduced to that problem with only polynomial overhead, this process could still be onerous in practice, and direct mapping may be more desirable As we demonstrated in section \ref{sub:wmis}, the diagonal $j=k$ elements of the QUBO can be controlled by modulating the relative phase applied to each mode. To implement an off-diagonal element, one must apply a controlled value of phase (not just $-1$) if $s_j=s_k=1$ and apply no phase otherwise, we can the superoperoperator in equation \ref{eq:Omega_nl_phase}.

By applying the superoperoperator in equation \ref{eq:Omega_nl_phase}, we apply an arbitrary phase of $\phi_{Q(jk)}$ in the case where two photons are present, which can only occur if there is one photon in each mode entering the overall circuit. Moreover, as discussed in section \ref{sub:HOM_constr}, the Hong-Ou-Mandel effect guarantees that the two photons will be in the same mode. We thus can implement an arbitrary off-diagonal QUBO element. Note that unlike solving an MIS problem, a QUBO solver is only possible if the pump mode is lossless.

However, we still need to guarantee that the systems starts in it's effective ground state. While it was a given that vacuum was the ground state in the maximum independent set case, since the inverted phase is effectively finding the \emph{minimum} independent set which is always the empty set for any graph, there is no similar guarantee for QUBOs, where elements can be positive or negative. There would be a number of ways to rectify this issue, a general strategy could be to perform a three parameter anneal, with the usual $\phi(\tau)$ and $c(\tau)$ parameters, but also with an additional term $\zeta(\tau)$, which can be used to tune the strength of the terms encoding a QUBO, such that each of them are performed with a normalised angle $\zeta(\tau)\phi_{Q(jk)}$. Such a protcol would start with $c(\tau=0)=\zeta(\tau=0)=0$ and $\phi(\tau=0)>0$ and monotonically tune to $c(\tau=1)=\phi(\tau=1)=0$ and $\zeta(\tau=1)>0$. Such a protocol would have the advantage of naturally biasing toward the $\ket{00...0}$ state\footnote{Since the definition of logical $0$ and $1$ is arbitrary, the QUBO can always be gauge transformed such that any desired state maps to $\ket{00...0}$.}, which is an additional useful feature since biasing techniques can serve an important role in hybrid quantum/classical algorithms \cite{Perdomo-Ortiz11,Duan2013a,chancellor17b,Grass19a,Callison2022hybrid}. We will not examine these protocols further in this work, but it provides an interesting avenue toward more general annealing.

\section{Exact solution for $\rho_{12}$ subject to Lossy Sum Frequency Generation\label{app:lossy_SFG_coherence}}

We wish to consider the dynamics of a system under the superoperator defined in equation \ref{eq:Omega_dl_SFG}. We are particularly interested in the action on the coherence between the $\ket{1}$ and $\ket{2}$ state as this matrix element is responsible for the Zeno effect which confines the system within the $\{\ket{0},\ket{1}\}$ manifold. In particular, assigning a phase to this element will lead to incoherent Zeno effects while removing this element will lead to incoherent Zeno effects, while combinations of the two will lead to intermediate effects. 

To make the time evolution of the system analytically tractable, we set $c=0$, but allow arbitrary $\gamma\ge 0$ and $\eta \ge 0$. From the definition of sum-frequency generation in equation \ref{eq:G_sfg} and the definition of single photon loss in \ref{eq:loss_lind} we see that the density matrix elements will obey:
\begin{align}
    \frac{\partial\rho_{12}}{\partial t}=i\sqrt{2}\gamma\rho_{1p} \\
    \frac{\partial\rho_{1p}}{\partial t}=i\sqrt{2}\gamma\rho_{12}-\frac{\eta}{2} \rho_{1p}
\end{align}
where $\rho_{12}=\sandwich{1,0_p}{\hat{\rho}}{2,0_p}$ and $\rho_{1p}=\sandwich{1,0_p}{\hat{\rho}}{0,1_p}$. Substituting these two equations into each other and eliminating $\rho_{1p}$, we obtain the differential equation for a damped harmonic oscillator
\begin{equation}
    0=\frac{\partial^2\rho_{12}}{\partial t^2}+\frac{\eta}{2} \frac{\partial\rho_{12}}{\partial t}+2\gamma^2\rho_{12}.
\end{equation}
The solution to this differential equation can be written as complex exponentials
\begin{equation}
    u_{\pm}(t)=\exp(\beta_{pm} t),
\end{equation}
where
\begin{equation}
    \beta_{\pm}=\frac{-\eta\pm\sqrt{\eta^2-32\gamma^2}}{4}.
\end{equation}
The damped Harmonic oscillator has three regimes of operation, the underdamped regime where $\beta_{\pm}$ are both complex, critical damping where $\sqrt{\eta^2-32\gamma^2}$ vanishes and therefore $\beta_{+}=\beta_{-}$, and the overdamped regime where $\beta_{\pm}$ are both real. The condition for critical damping is that
\begin{equation}
    \eta=4\sqrt{2}\gamma
\end{equation}

\begin{figure}
    \centering
    \includegraphics[width=0.75\linewidth]{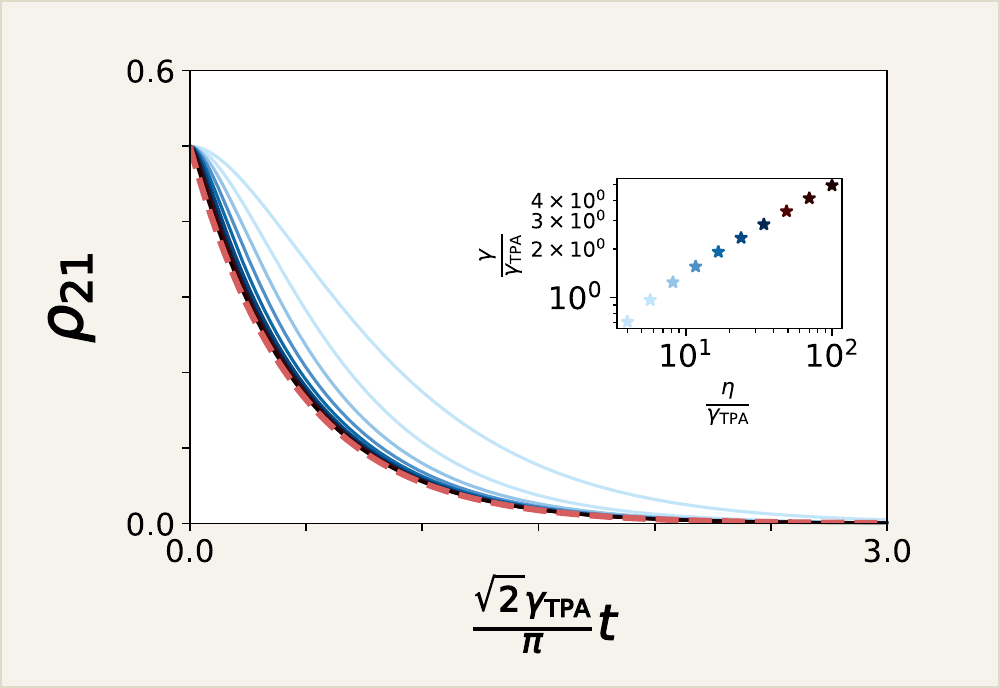}
    \caption{Plot of $\rho_{12}=\sandwich{1,0_p}{\hat{\rho}}{2,0_p}$ versus time for time evolution with the superoperator $\Omega_\mathrm{dl,SFG}(c=0,\gamma,\eta,t)\left[\hat{\rho}_0\right]$ from equation \ref{eq:Omega_dl_SFG}. The system is initialized in $\hat{\rho}_0=\frac{1}{2}(\ket{1}+\ket{2})(\bra{1}+\bra{2})$ with the pump mode empty for $c=0$ and varying values of $\eta$ for a fixed $\gamma_\mathrm{TPA}$ (see equation \ref{eq:gamma_TPA}) ranging from critically damped ($\eta=4\sqrt{2}\gamma$, $\eta=4\gamma_\mathrm{TPA}$), to strongly overdamped ($\eta=100 \gamma_\mathrm{TPA}$. The inset shows the values of $\gamma$ and $\eta$ using the same color scheme. The red dashed line shows the Markovian two-photon absorption with a loss rate of $\gamma_\mathrm{TPA}$ for comparison.}
    \label{fig:rho_12_markov_approach}
\end{figure}

The underdamped regime will feature oscillations in which the density matrix element $\rho_{12}$ will become negative and therefore this regime can include some phase based Zeno effects. Explicitly, if we set our initial conditions as $\rho_{12}(0)$ and specify that the pump mode will initially be empty and therefore $\frac{\partial\rho_{12}}{\partial t}\big|_{t=0}=0$, the underdamped solution can be written as
\begin{align}
    \rho_{12}(t)=\nonumber \\\rho_{12}(0)\left[\cos\left(\sqrt{2\gamma^2-\frac{\eta^2}{16}}t\right)+\frac{\eta}{4}\sin\left(\sqrt{2\gamma^2-\frac{\eta^2}{16}}t\right)\right]\exp\left(-\frac{\eta}{4}t \right).
\end{align}
Right at the critical damping point, the solution is
\begin{equation}
    \rho_{12}(t)=\rho_{12}(0)(1+\frac{\eta}{4}t)\exp\left(-\frac{\eta}{4}t \right).
\end{equation}
and the overdamped solution is
\begin{align}
    \rho_{12}(t)=\rho_{12}(0)\Bigg[\frac{4+\eta}{8}\exp\left(-\left(\frac{\eta}{4}-\sqrt{\frac{\eta^2}{16}-2\gamma^2}\right)t \right)+\nonumber \\\frac{4-\eta}{8}\exp\left(-\left(\frac{\eta}{4}+\sqrt{\frac{\eta^2}{16}-2\gamma^2}\right)t \right) \Bigg].
\end{align}
To study an interpolation between coherent and incoherent Zeno effects we should stay in the range between under and critical damping so $0\le\eta\le 4\sqrt{2}\gamma$. At the critically damped point, there is no more phase inversion so a Zeno effect will be entirely incoherent. The combined effect of the pump mode and the loss is however still highly non-Markovian, unlike the two-photon absorption Lindbladian in equation \ref{eq:tpa_lind} which is Markovian by construction. The overdamped regime can be seen to recover the Markovian limit as $\frac{\eta}{\gamma}\rightarrow \infty$, this can be seen by noting that the second term in the solution will vanish very quickly, leaving a single exponential decay characteristic of a Markovian bath. If we Taylor expand the coefficient of the slower decaying term, the one which will be realized in an effective two-photon absorption model we find that
\begin{equation}
    \gamma_\mathrm{TPA}=\frac{\eta}{4}-\sqrt{\frac{\eta^2}{16}-2\gamma^2}=\frac{4 \gamma^2}{\eta}+O(\gamma^4) \label{eq:gamma_TPA}
\end{equation}
on the other hand, when two-photon absorption of the form given in equation \ref{eq:tpa_lind} is applied, the $\rho_{12}$ density matrix would decay with a rate of $\gamma$. By setting the two exponents equal, we can see that in the large $\eta$ limit, to match a two-photon absorption rate of $\gamma_\mathrm{TPA}$, with sum-frequency generation and loss, we should set
\begin{equation}
    \gamma\approx\frac{\sqrt{\eta}}{2}\gamma_\mathrm{TPA}.
\end{equation}
If we set constant $\gamma_\mathrm{TPA}$ in equation \ref{eq:gamma_TPA}, and solve for $\gamma$ at given $\eta$ values, we can see the lossy sum-frequency generation model approach a Markovian two-photon absorption description with increasing $\eta$, this behavior is shown in figure \ref{fig:rho_12_markov_approach}.

\section{Fully Connected Delay Protocol \label{app:delay_reorder}}

\begin{figure}
    \centering
    \includegraphics[width=0.5\linewidth]{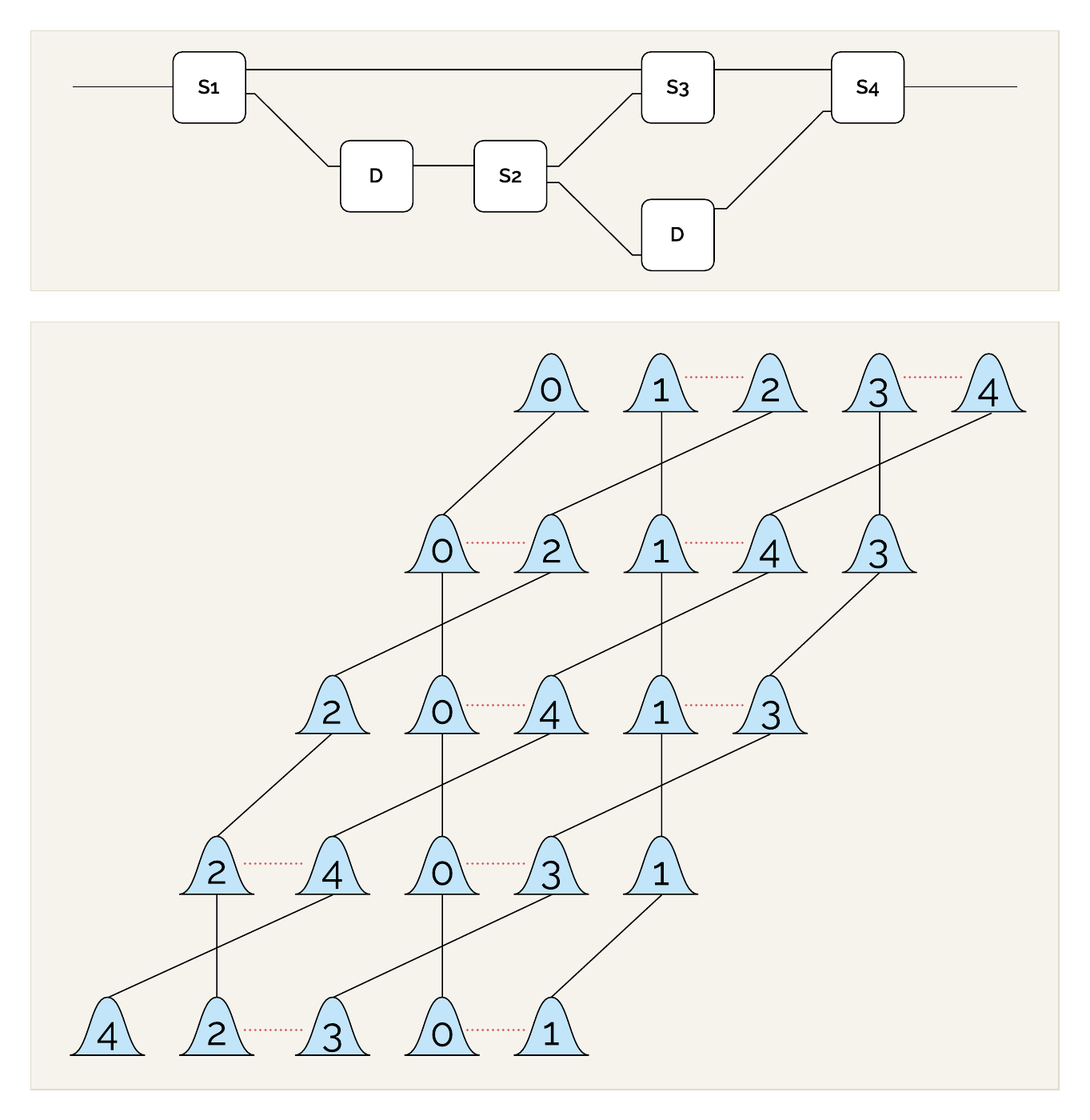}
    \caption{ An example of $n=5$ rounds of algorithm \ref{alg:time_bin_shuffle} to interact $n=5$ time bins. Dashed lines illustrate that all will be able to interact even accounting for the fact that a time bin can only interact with one of its neighbors per round. Optical switching and delays to implement the necessary shifts to implement the protocol.}
    \label{fig:shuffle_protocol}
\end{figure}

For highly connected graphs it is necessary to allow every pair of time bins an opportunity to interact via a constraint. To achieve this, in the middle of a train of time bins we apply delays of $2D$ alternatively on the odd and even numbered time bins in a sequence of time bins. Special actions need to be taken at the end points, for example when the last bin (labeled 0 in our convention where we count from the back) is delayed we only delay it by $D$ so that it remains adjacent to the next time bin. Furthermore if the second bin is delayed by $2D$, the first bin in the train must be delayed but $D$ to remain adjacent. Pseudocode for this protocol is shown as algorithm \ref{alg:time_bin_shuffle}, and it is visualized in figure \ref{fig:shuffle_protocol}. 

\begin{algorithm}
    \caption{Algorithm for high-level time bin shifting to such that all pairs of time bins are able to interact. Assumes $n$ modes. Two subroutines are used $ApplyDrivingAndInteractions$ performs the relevant physical processing based on the mode order while $ApplyDelays$ applies delays of $0$, $D$, or $2D$ to each mode.}
    \label{alg:time_bin_shuffle}
    \begin{algorithmic}
        \State $mode\_nums \gets [0,1,2,...n-1]$ \Comment{one dimensional array to track locations of modes}
        \For{$i\_shuffle$ from $0$ to $n-1$}
        \State $delay\_list \gets []$ \Comment{empty list for delays}
        \State $mode\_num\_new \gets [0]\times n$ \Comment{all zeros}
            \For{$i\_mode$ from $0$ to $n$}
                \If{$\mod_2(i\_mode)=\mod_2(i\_shuffle)$}
                    \If{$i\_shuffle=0$}
                        \State $delay\_list \gets delay\_list+[D]$ \Comment{Only delay by $D$ if first entry}
                        \State $mode\_nums\_new[0] \gets mode\_nums[0]$
                    \Else
                        \State $delay\_list \gets delay\_list+[2D]$
                        \State $mode\_nums\_new[i\_shuffle-1] \gets mode\_nums[i\_shuffle]$
                    \EndIf
                \ElsIf{$i\_shuffle=n-1$}
                    \State $delay\_list \gets delay\_list+[D]$ \Comment{Delay last entry as special case}
                    \State $mode\_nums\_new[n-1] \gets mode\_nums[n-1]$
                \Else
                    \State $delay\_list \gets delay\_list+[0]$ 
                    \State $mode\_nums\_new[i\_shuffle+1] \gets mode\_nums[i\_shuffle]$
                \EndIf
            \EndFor
            \State $ApplyDrivingAndIneteractions(mode\_nums)$ 
            \State $ApplyDelays(delay\_list)$
            \State $ mode\_nums \gets mode\_nums\_new$
        \EndFor
    \end{algorithmic}
\end{algorithm}

Because this protocol only involves delays of $0$, $D$, or $2D$, the shifting can be implemented using the array of switches and delays depicted in the upper portion of figure \ref{fig:shuffle_protocol}. From the inset we can see that even if we only do interactions in one direction (depicted by dashed lines), each time bin will have an opportunity to interact with all $n-1$ other time bins.

\section{Detailed Numerical Methods \label{app:detailed_num}}

The numerical results presented in this manuscript were achieved by using a full state-space representation of photonic systems and leveraging the fact that as linear equations, Masters equations can be solved by exponentiating a matrix representation of the superoperator. We use two functions available through SciPy, \textbf{expm} and \textbf{expm\_multiply} to achieve this. The difference between these two functions is that \textbf{expm} creates a (dense) matrix representing the exponentiation of a matrix which can be multipled by arbitrary state vectors while \textbf{expm\_multiply} computes the action of an exponentiated sparse matrix on a state vector directly, without ever constructing a dense matrix.

In practice, to achieve runtimes which are accessible without HPC resources, we have optimized the code to reduce computational overheads. The main difficulty we found was that in some cases we need to simulate a large number of cycles with slightly changed phases and driving in each. These large number of cycles suggest that pre-computing as many quantities as possible to avoid (relatively) time consuming calculations in the inner loop is the best way to improve performance. The two operations which we wish to avoid in this setting are tensor products and matrix exponentiation. For constraint terms, this can be achieved very simply by computing the evolution superoperator once at the beginning for each edge of the graph, since these remain constant at each cycle. The phase and driving operations however present a challenge since they vary dynamically throughout the anneal, we apply different strategies for each.

\subsection*{Elementwise Phase Application}

The superoperator for applying phases has a diagonal structure in the sense that in the computational basis, it maps each density matrix element to itself with a possible phase. Moreover, if a phase of $\phi$ is applied only to a single qubit variable, then each density matrix element will only ever get multipled by $1$, $\exp(i \phi)$, or $\exp(-i\phi)$. By pre-computing which elements get which phase for each phase superoperator, we can use this information to construct a superoperator for an arbitrary $\phi$ without having to perform any tensor products or matrix exponentiations, only two exponentiations of scalar variables.

\subsection*{Precomputed Binary Decomposition for Driving}

The driving superoperators unfortunately do not have the simple structure of the phase superoperators, especially if we consider an imperfect Zeno blockade which allows the system to leave the $\{\ket{0},\ket{1} \}$ subspace. While the tensor products used to construct the generators can simply be precomputed and used, such a strategy would still leave a computationally expensive matrix exponentiation to be performed in the inner loop. To avoid matrix exponentiation in the inner loop, we use the property that 
\begin{equation}
    \exp\left(t\mathcal{G}\right)=\exp\left(\left(\sum_jt_j\right)\mathcal{G}\right)=\prod_j\exp\left(t_j\mathcal{G}\right)
\end{equation}
where $t_j$ are aribitrary scalar values such that $\sum_jt_j=t$ and $\mathcal{G}$ is an arbitrarily generating matrix. This relationship can be understood physically if we imagine $\mathcal{G}$ describing a differential equation and $t_j$ describing time intervals of different duration. Clearly applying time evolution for a number of intervals adding up to at total time is equivalent to directly evolving for the total. 

Using this property we can decompose 
\begin{equation}
    \exp\left(t\mathcal{G}\right)=\prod_j\exp\left(\frac{t_\mathrm{max}b_j}{2^j}\mathcal{G}\right)
\end{equation}
where $ b_j\in \{0,1\}$ comprise a binary decomposition such that 
\begin{equation}
    t=\sum^{m-1}_{j=0}\frac{b_jt_\mathrm{max}}{2^j}+O\left(\frac{t_\mathrm{max}}{2^m}\right).
\end{equation}
We therefore can compute a highly accurate approximation of the time evolution superoperator for any time less than $2t_\mathrm{max}$ by precomputing $m$ matrix exponentials corresponding to each runtime, and only performing matrix multiplication within the inner loop. Given the exponential accuracy increase with $m$ high accuracy can still be obtained for a modest value of $m=30$. When the runtime is dominated by runs where $m\ll n_\mathrm{cycle}$, the number of computationally-expensive matrix exponentiations is greatly reduced.

\bibliographystyle{unsrt}
\bibliography{references}  % bibtex entries in separate file for convenience

\end{document}